\documentclass[a4paper,11pt]{article}

\usepackage{graphicx}  
\usepackage{dcolumn}   
\usepackage{bm,relsize}        
\usepackage{amssymb, amsmath}
\usepackage{textcomp}
\usepackage{wasysym}
\usepackage{slashed}
\usepackage{caption, subcaption}
\usepackage{multirow}
\usepackage{gensymb}
\usepackage{subcaption}
\usepackage{colortbl}
\usepackage{booktabs} 
\usepackage{tabularx}
\definecolor{headergray}{gray}{0.9}
\definecolor{rowgray}{gray}{0.97}

\usepackage{cancel}

\usepackage[nolist, nohyperlinks]{acronym}

\usepackage{lipsum, color}
\usepackage[dvipsnames,svgnames,table]{xcolor}
\usepackage{braket}
\usepackage{jheppub} 
\usepackage{booktabs}
\usepackage{pdflscape}

\usepackage{booktabs}
\usepackage{tabularx}

\usepackage{mathrsfs}
\usepackage{bm,amssymb,slashed,graphicx,multirow,soul,mathtools,xspace,array,tikz,amsmath, gensymb}
\usepackage[compat=1.1.0]{tikz-feynman}
\usetikzlibrary{patterns}
\usepackage{siunitx} 
\usepackage{float}   
\usepackage{cancel}
\allowdisplaybreaks
\usepackage{ bbold }
\usepackage{hyperref}
\usepackage{colortbl}
\usepackage{tcolorbox}
\usepackage{tikz}
\usetikzlibrary{shapes.geometric}
\usetikzlibrary{shapes.misc}
\usetikzlibrary{patterns.meta}
\usepackage{comment}

\usepackage[capitalise, english, nosort]{cleveref}

\definecolor{nicered}{rgb}{0.7,0.1,0.1}
\definecolor{nicegreen}{rgb}{0.1,0.5,0.1}
\definecolor{violet}{rgb}{0.7,0.3,0.3}
\hypersetup{colorlinks,citecolor= nicegreen,linkcolor= nicered}

\newcommand{\lp}{\left(}
\newcommand{\rp}{\right)}

\newcommand{\g}{\gamma}
\newcommand{\be}{\begin{equation}}
\newcommand{\ee}{\end{equation}}

\newcommand{\eventname}{KM3-230213A}
\newcommand{\kmn}{KM3NeT}
\newcommand{\ic}{IceCube}

\newcommand{\TeV}{\si{\tera\electronvolt}}
\newcommand{\GeV}{\si{\giga\electronvolt}}
\newcommand{\PeV}{\si{\peta\electronvolt}}

\newcommand{\dd}[0]{\mathrm{d}}

\newcommand{\beq}{\begin{equation} }
\newcommand{\eeq}{\end{equation}} 
\newcommand{\bi}{\begin{itemize} }
\newcommand{\ei}{\end{itemize} }

\newcommand{\deriv}[2]{\frac{\partial #1}{\partial #2}}

\definecolor{Red}{rgb}{1.,0.,0.}
\definecolor{Grn}{rgb}{0.,0.75,0.}
\definecolor{Blu}{rgb}{0.,0.,1.}
\definecolor{Pink}{rgb}{1,0.08,0.58}

\DeclareMathOperator{\erf}{Erf}

\usepackage[T1]{fontenc} 

\usepackage{amsmath,amssymb,epsfig,color,slashed}
\allowdisplaybreaks  

\newcommand{\gcm}{{\rm g}/{\rm cm}^3} 

\newcommand{\yw}{y_{\rm w}}
\newcommand{\py}{\mathcal{P}(\yw)}

\definecolor{verdino}{rgb}{0.66, 0.89, 0.63}
\definecolor{coral}{RGB}{255,127,80}
\definecolor{cadetblue}{RGB}{95,158,160}
\definecolor{blueviolet}{RGB}{138,43,226}

\newcommand{\eCC}[1]{
  \resizebox{#1}{!}{
        \begin{tikzpicture}[baseline]
          \node at (0,0) [draw=none, fill=cadetblue!50, minimum size=2.2cm, thick] {};

          \node at (.3cm, .3cm) [circle, draw=none, minimum size = .6cm, fill=blueviolet!65,very thick] (cc) {};

          \node at (.2cm, .2cm) [circle, draw, minimum size = .6cm ,color=blueviolet,fill=blueviolet,very thick] (cc1) {};

        \end{tikzpicture}
      }
      }
\newcommand{\cascade}[1]{
  \resizebox{#1}{!}{
        \begin{tikzpicture}[baseline]
          \node at (0,0) [draw=none, fill=cadetblue!50, minimum size=2.2cm, thick] {};

          \node at (.2cm,.2cm) [circle, draw=none, minimum size =.6cm,fill=blueviolet!65,very thick] (cc) {};

        \end{tikzpicture}
      }
    }

\newcommand{\muontracks}[1]{
  \resizebox{#1}{!}{
    \begin{tikzpicture}[baseline]
        \node at (0,0) [draw=none, fill=cadetblue!50, minimum size=2.2cm, thick] {};
      \node at (1.5cm,0) [draw=none, fill=coral!30, minimum width=.8cm, minimum height=2.2cm] {};
      
      \node at (1.8cm, .7cm)  (s) {};
      \node at (1.2cm, .4cm)  [inner sep=0pt]  (m) {};
      \node at (-.5cm,-.45cm) (e) {};
      \draw[dashed, color=blueviolet] (s) -- (m);
      \draw[color=blueviolet] (m) -- (e);
      \node at (1.8cm, .4cm)  (s1) {};
      \node at (.8cm, -.1cm)  [inner sep=0pt]  (m1) {};
      \node at (-.5cm,-.75cm) (e1) {};
      \draw[ dashed,color=blueviolet] (s1) -- (m1);
      \draw[ dash phase=0pt,color=blueviolet] (m1) -- (e1);
      \node at (.2cm, .65cm) [circle, draw,color=blueviolet!65,fill=blueviolet!65,very thick] (cc2) {};
      \node at (-.8cm,.15cm) (e2) {};
      \draw [color=blueviolet] (cc2) -- (e2);
    \end{tikzpicture}
  }
}
\newcommand{\tautracks}[1]{
  \resizebox{#1}{!}{
    \begin{tikzpicture}[baseline]
      \node at (0,0) [draw=none, fill=cadetblue!50, minimum size=2.2cm, thick] {};
      \node at (1.5cm,0) [draw=none, fill=coral!30, minimum width=.8cm, minimum height=2.2cm] {};
      
      \node at (1.8cm, .8cm)  (s) {};
      \node at (1.2cm, .5cm)  [inner sep=0pt]  (m) {};
      \node at (-.5cm,-.35cm) (e) {};
      \draw[ dashed,color=blueviolet] (s) -- (m);
      \draw [color=blueviolet](m) -- (e);
      \node at (1.8cm, .5cm)  (s1) {};
      \node at (.8cm, 0)  [inner sep=0pt]  (m1) {};
      \node at (-.5cm,-.65cm) (e1) {};
      \draw[ dashed,color=blueviolet] (s1) -- (m1);
      \draw[ dash phase=0pt,color=blueviolet] (m1) -- (e1);
      \node at (.2cm, .75cm) [circle, draw,color=blueviolet!65,fill=blueviolet!65,very thick] (cc2) {};
      \node at (-.8cm,.25cm) (e2) {};
      \draw[color=blueviolet] (cc2) -- (e2);
      \node at (.2cm, .49cm)  (e3) {};
      \node at (-.75cm,.015cm) [circle, draw,color=blueviolet!65,fill=blueviolet!65,very thick](cc3) {};
      \draw[color=blueviolet] (cc3) -- (e3);
      \node at (.7cm, -.45cm) [circle, draw, minimum size = 1mm ,color=blueviolet!65,fill=blueviolet!65,very thick] (cc) {};
      \node at (0, -.8cm) [circle, draw, minimum size = 1mm ,color=blueviolet!65,fill=blueviolet!65,very thick] (cc1) {};
      \draw[color=blueviolet] (cc) -- (cc1);
    \end{tikzpicture}
  }
}

\tikzset{cross/.style={cross out, draw=black, minimum size=2*(#1-\pgflinewidth), inner sep=0pt, outer sep=0pt},
  cross/.default={1pt}}

\newcommand{\inside}[1]{
  \resizebox{#1}{!}{
    \begin{tikzpicture}[anchor=center,baseline]
      \draw [fill=cadetblue, color=cadetblue, line width =.4mm] (0,0) circle [radius=1.22 cm];
      \node at (.5cm, .5cm) (ev) {};
      \draw (ev) node[cross, blueviolet, thick, minimum size=.2cm, rotate=45]{};
      \node at (2.2cm,2.2cm) (nu) {};
      \draw [color=blueviolet, dashed] (nu) -- (ev);
    \end{tikzpicture}
  }
}

\newcommand{\outside}[1]{
  \resizebox{#1}{!}{
    \begin{tikzpicture}[anchor=center,baseline]
      \draw[rotate=45, dashed, color=cadetblue, pattern={Lines[distance=1mm, angle=45,line width=.2mm]}, pattern color=cadetblue] (0,0) circle [x radius=.1cm, y radius=1.25cm];
      \draw[rotate=45, dashed, color=cadetblue, pattern={Lines[distance=1mm, angle=-30,line width=.2mm]}, pattern color=cadetblue] (0,0) circle [x radius=.1cm, y radius=1.25cm];
      \draw [rotate=135, dashed, color=cadetblue] (1.25,0) arc(0:180:1.25cm and .1cm);
      \draw [rotate=-45, color=cadetblue, line width =.4mm] (1.25cm,0) arc(0:180:1.25cm);
      \draw [rotate=135, color=cadetblue, line width =.4mm] (1.25cm,0) arc(0:180:1.25cm);

      \begin{scope}
        \clip [rotate=-45] (1.23cm,0) arc(0:180:1.23cm) arc(180:0:1.23cm and .1cm);
        \fill [pattern={Lines[distance=1.5mm, angle=45, line width=.2mm]}, pattern color=cadetblue!40] (-2,-2) rectangle (2,2);
      \end{scope}

      \node at (1.8cm, 1.8cm)   [inner sep=0pt] (mu) {};
      \node at (.9cm, .9cm) (ev) {};
      \draw (ev) node[cross, blueviolet, thick, minimum size=.2cm, rotate=45]{};
      \node at (2.5cm,2.5cm) (nu) {};
      \draw [color=blueviolet, dashed] (nu) -- (mu);
      \draw [color=blueviolet] (mu) -- (ev);
      \node at (1.cm, 1.8cm)   [inner sep=0pt] (mu1) {};
      \node at (.4cm, 1.2cm) (ev1) {};
      \draw (ev1) node[cross, blueviolet, thick, minimum size=.2cm, rotate=60]{};
      \node at (1.7cm,2.5cm) (nu1) {};
      \draw [color=blueviolet, dashed] (nu1) -- (mu1);
      \draw [color=blueviolet] (mu1) -- (ev1);
      \node at (1.15cm, -.4cm) (ev2) {};
      \draw (ev2) node[cross, blueviolet, thick, minimum size=.2cm, rotate=60]{};
      \node at (2.55cm,1cm) (nu2) {};
      \node at (1.85cm, .3cm)   [inner sep=0pt] (mu2) {};
      \draw [color=blueviolet, dashed] (nu2) -- (mu2);
      \draw [color=blueviolet] (mu2) -- (ev2);
    \end{tikzpicture}
  }
}

\begin{document} 

\begin{acronym}
\acro{IC}{IceCube}
\end{acronym}


\title{The soft volume of ultra-high energy neutrinos experiments}

\author[a]{Stefano Palmisano,}
\author[a]{Diego Redigolo,}
\author[b]{Michele Tammaro,}
\author[a]{Andrea Tesi}
\affiliation[a]{INFN Sezione di Firenze, Via G. Sansone 1, I-50019 Sesto Fiorentino, Italy}
\affiliation[b]{Galileo Galilei Institute for Theoretical Physics, INFN, Largo Enrico Fermi 2, I-50125 Firenze, Italy}

\emailAdd{stefano.palmisano@fi.infn.it}
\emailAdd{diego.redigolo@fi.infn.it}
\emailAdd{michele.tammaro@fi.infn.it}
\emailAdd{andrea.tesi@fi.infn.it}

\date{\today}

\preprint{}

\abstract{We develop a semi-analytical framework to map ultra-high-energy neutrino fluxes onto event rates at neutrino telescopes. The formulation is based on the Boltzmann equation for the distribution of secondary muons produced by neutrino interactions in matter, and naturally accounts for the effective target volume relevant for through-going tracks. This ``soft volume'' is controlled by muon propagation and stochastic energy losses, and can be significantly larger than the instrumented detector volume. Exploiting the dominance of soft energy losses, we derive a controlled second-order expansion of the collision operator, reducing the transport problem to a drift-diffusion equation in energy space, with rare hard scatterings treated perturbatively. The resulting master formula provides a fast alternative to full Monte Carlo simulations while retaining a direct connection to the microscopic muon energy-loss processes. We apply the formalism to IceCube through-going muon data, marginalizing over theoretical uncertainties in the transport coefficients, and obtain a diffuse-flux fit compatible with the experimental result. We also revisit the interpretation of the ultra-high-energy track event reported by KM3NeT, assessing its consistency with IceCube non-observation in the same energy range. Our results provide a first-principles bridge between neutrino-flux models and track-event observables, with direct applications to precision neutrino astronomy and searches for physics beyond the Standard Model.
}

\maketitle

\section{Introduction}
\label{sec:intro}

In this work, we develop an ab-initio semi-analytical mapping between the neutrino flux impinging on a detector and the corresponding event rates observed at neutrino telescopes, taking a step forward from Ref.~\cite{Palmisano:2025abd}. Such a mapping is a necessary ingredient for a systematic interpretation of ultra-high energy (UHE) neutrino data, and in particular for robust searches for new-physics signals in neutrino telescope observations.

\begin{figure}
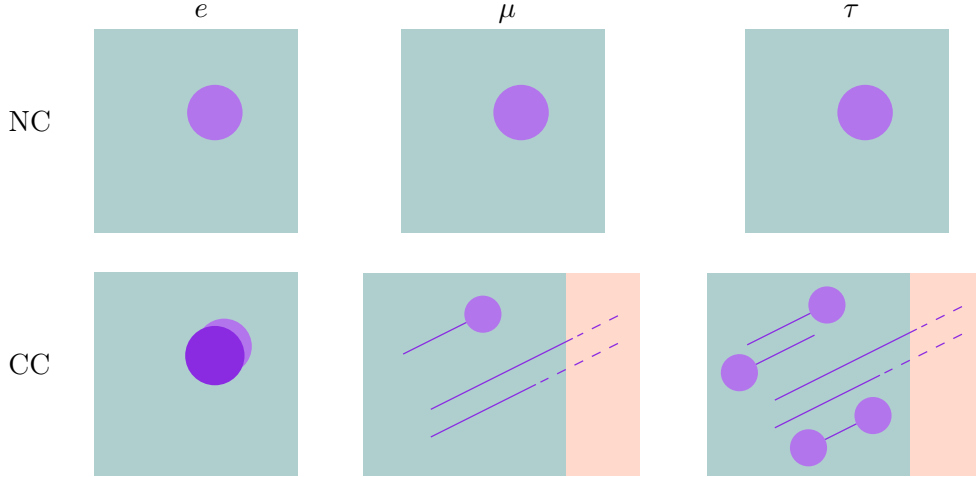

  \centering
  \begin{tabular}{lccc}
    & $e$ & $\mu$ & $\tau$ \\
    NC
    &
      \cascade{3cm}
    &
        \cascade{3cm}
    &
      \cascade{3cm}
    \\
    &&&
    \\
    CC
    &
    \eCC{3cm}
    &
    \muontracks{4cm}
    &
    \tautracks{4cm}
    \\
    \end{tabular}
    \caption{Event categories at neutrino telescopes for each neutrino flavor (columns), initiated by NC or CC interactions (rows). Violet blobs represent localized energy depositions, from hadrons or electrons, whereas solid lines represent either muons or taus. They can be produced outside the detector (represented by the green-filled square), which naturally leads to the identification of a \emph{soft volume} (orange rectangle).\label{fig:eventshapes}}
\end{figure}

We begin by recalling the event topologies at neutrino telescopes, a simple but important point that plays a central role throughout this work. Signals at neutrino telescopes are recorded through the Cherenkov light produced by secondary charged particles and detected by photomultipliers (PMTs). For instance, IceCube consists of an array of PMTs, known as Digital Optical Modules, deployed along 86 strings~\cite{IceCube:2008qbc}. Events can be characterized according to their topology, see \cref{fig:eventshapes}, and mainly fall into two categories: \emph{track events} and \emph{cascade events}. The former are characterized by an elongated pattern of activated PMTs, indicating the passage of a charged lepton with a sufficiently long propagation length~\cite{IceCube:2023agq}. The latter are instead characterized by more localized, approximately spherical light patterns, associated with hadronic showers or electromagnetic (EM) showers with short propagation lengths~\cite{IceCube:2020acn}.

The separation into these two categories can be understood within the Standard Model (SM) by following how events are generated. Assuming an approximately flavor-democratic neutrino flux, as expected accounting for neutrino oscillations over astrophysical distances~\cite{Athar:2000yw}, neutrinos can scatter in the Earth or inside the detector through neutral-current (NC) or charged-current (CC) interactions. In NC interactions, the visible final state consists of the hadronic shower generated by the scattering. Since hadrons interact strongly and electromagnetically with matter, these showers deposit most of their energy over very short distances compared to the typical sizes and resolutions of neutrino telescopes. Thus, NC events of all three flavors are detectable as contained cascade events only when the scattering occurs inside, or sufficiently close to, the detector volume.

In CC interactions, the visible final state consists of both the hadronic shower at the interaction point and the corresponding charged lepton: an electron, a muon, or a tau. Electrons initiate EM showers and therefore behave similarly to hadronic showers from the point of view of the event topology, depositing their energy over short distances. It follows that electron-neutrino CC events also fall into the cascade category at neutrino telescopes.
Conversely, muons can propagate over large distances before losing most of their energy through radiative processes. Muon-neutrino CC interactions therefore predominantly contribute to the track category. These events represent a large fraction of the expected event rate. Indeed, considering the rough estimate $\sigma_{\rm CC}/\sigma_{\rm NC}\approx3$, CC-induced events constitute about $75\%$ of the total interaction rate. Assuming a flavor-democratic flux, muon-neutrino CC interactions then account for roughly one third of the CC rate, namely about $25\%$ of the total neutrino interaction rate. Interestingly, this approximation underestimates the observed muon-track rate, since muons can be produced far outside the detector, up to distances of order of a few kilometers, and still retain enough energy to leave a track in the instrumented volume. 

The central observation is, therefore, that for track-like events the relevant target volume is not fixed solely by the instrumented detector geometry. It is instead a dynamical quantity, which we dub \emph{soft volume}, controlled by the propagation and energy-loss history of the muon before it reaches the detector. This effect is absent, or strongly suppressed, for cascade-like events, whose visible final states are localized near the interaction point. Very roughly, the rate of events in each category can be estimated as
\begin{equation}
\label{eq:Nexpected:Sketch}
\begin{split}
\langle N_i \rangle
    &\propto {\color{blueviolet} V_{\mathrm{detector}}} \oplus {\color{coral} V_{\mathrm{soft}}}\ , \\
    &\propto \left(\text{detector area}\right)
    \times
    \left[
    {\color{blueviolet}\left(\text{detector radius}\right)}
    \oplus
    {\color{coral}\left(\text{energy-loss range of } X_i\right)}
    \right]\ ,
\end{split}
\end{equation}
where the index $i$ labels the event category, with corresponding final state $X_i$. The last quantity denotes the typical length scale over which $X_i$ can propagate before losing a significant fraction of its energy, and identifies the soft volume. For muons, this length scale is typically of the order of the kilometer. The soft-volume enhancement is, in principle, measurable in data by comparison between observed cascade and track events, as well as within the track sample itself by comparing different IceCube event selections. 

A separate discussion is required for taus, whose topology depends on both energy and containment. For energies well below $10~\mathrm{PeV}$, taus have a decay length that is short compared to the detector size as shown in \cref{fig:eventshapes}, and their decay products typically appear as cascade-like activity unless the displacement between the production and decay vertices is resolvable, in which case the event gives rise to a double-bang signature~\cite{IceCube:2020fpi,IceCube:2024nhk}. Depending on containment, tau events can also appear with so-called \emph{lollipop} or \emph{inverted-lollipop} topologies~\cite{Beacom:2003nh}, while muonic tau decays, $\tau\to\mu\nu_\tau\bar{\nu}_\mu$, can produce mixed cascade-plus-track events or tracks with characteristic changes in light yield~\cite{DeYoung:2006fg}. At higher energies, above $100~\mathrm{PeV}$, the tau decay length exceeds the detector size, but the small tau energy loss per unit length makes the detection and reconstruction of tau tracks challenging. We show in \cref{sec:TauInducedMuons} that the contribution of tau leptons to the track sample of IceCube recorded data is negligible. Tau-induced cascades and mixed topologies, however, may also benefit from a soft-volume enhancement when taus are produced outside the instrumented volume and subsequently decay inside the detector. Quantifying this contribution is left for future work.

In the following, we focus instead on the muon case, where the soft-volume enhancement is already relevant for present data. The characterization of this soft volume was initiated in Ref.~\cite{Palmisano:2025abd}, where we developed a first-principles semi-analytical framework based on the Boltzmann equation for the muon distribution function, $f_\mu(x,p,t)$, sourced by an astrophysical neutrino flux. This formulation follows the full distribution function and therefore explicitly conserves the particle number, 
\beq\label{eq:N_from_fmu}
N_\mu \propto \int d^3x\, d^3p\, f_\mu(x,p,t)\,,
\eeq
in contrast to more traditional approaches~\cite{Gaisser:2016uoy}, allowing one to derive the expected flux directly in terms of the measured muon energy.

In this work, we streamline and extend this formalism by including higher-order contributions to stochastic muon energy losses in matter. The Boltzmann equation is, in general, an integro-differential equation, since the unknown muon distribution appears under the integral sign in the collision term. We exploit the fact that this term is dominated by \emph{soft} collisions, which we treat perturbatively in a small-energy-loss expansion, while \emph{hard} collisions are suppressed by the smaller rate of the hard processes. We show that the soft collisions can be solved semi-analytically at second order, leading to a drift-diffusion equation in the energy variable. Hard scatterings can be in principle incorporated as perturbative corrections. This approach enables a detailed and fast treatment of track events at neutrino telescopes without the use of intensive Monte Carlo (MC) simulations~\cite{Koehne:2013gpa}. More precisely, we perform microscopic simulations of muon energy losses to inform the prior range of the parameters controlling drift and diffusion. This procedure can be done once and for all and can be substituted by any suitable means to estimate the prior range of the parameters.

For this initial application, we adopt an idealized detector model in which every muon crossing the projected detector area is counted. We therefore do not include the full IceCube event selection, angular acceptance, reconstruction efficiency, or cuts on the deposited energy. Our prediction should consequently be interpreted as a geometric through-going muon rate rather than a detector-level event rate. Assuming a single power-law diffuse flux, as defined in \cref{eq:diffuse-definition}, we find
\begin{equation}
\gamma=2.38^{+0.11}_{-0.09}\,, \qquad \phi_0=0.63^{+0.14}_{-0.13}\times10^{-18}\,\rm{GeV}^{-1}\,\sec^{-1}\,{\rm cm}^{-2}\,{\rm sr}^{-1}\, .    
\end{equation}
The inferred spectral slope is in good agreement with the IceCube result, whereas the normalization is smaller. This is expected: for a fixed incident flux, the idealized perfect-detector approximation overestimates the number of accepted muons, and the fit compensates for this by preferring a lower flux normalization. The ratio between our inferred normalization and the IceCube best-fit value can therefore be interpreted as an effective acceptance-efficiency factor for the through-going analysis,
\begin{equation}\label{eq:ICeff}
\epsilon_{\rm IC\text{-}TG}
\equiv
\frac{\phi_0}{\phi_0^{\rm IC}}
\simeq 0.45\,.
\end{equation}
This factor collectively accounts for the event-selection, geometric-acceptance, reconstruction, and deposited-energy effects that are absent from our idealized analysis.

Second, we re-examine the ultra-high-energy neutrino event observed by \kmn{}~\cite{KM3NeT:2025npi}, which was identified as a single muon track produced outside the detector in a neutrino charged-current interaction, namely a \emph{through-going} muon track. This event has attracted considerable attention as a possible window onto exotic neutrino fluxes associated, for instance, with dark matter or early-Universe phenomena~\cite{Airoldi:2025opo,Su:2025qzt,Borah:2025igh,Yamamoto:2026ybj}. As pointed out in Refs.~\cite{Palmisano:2025abd,titans}, its observation is in tension with the absence of events in the same energy range at \ic{}, whose instrumented-volume exposure exceeds that of \kmn{} by more than an order of magnitude. We reassess this tension while consistently accounting for theoretical uncertainties in the treatment of muon energy losses. Quantifying the discrepancy through a Bayes factor (BF; see \cref{sec:FitToData} and Ref.~\cite{Palmisano:2025abd}), we find $\mathrm{BF}=18$, corresponding to a \emph{strong} tension according to the standard Jeffreys scale~\cite{Jeffrey}. Setting aside the comparison with \ic{}, we also determine what would be required to obtain one such event at \kmn{} within a Standard Model interpretation, and show that this would require an exceedingly large neutrino charged-current cross section~\cite{Bertolez-Martinez:2026bzj} which would quickly violate the Froissart bound~\cite{Froissart:1961ux} within the SM and seems difficult to explain with beyond the Standard Model physics without being excluded by complementary probes of the electroweak scale.

The rest of the work is organized as follows. In \cref{sec:setup}, we set up the Boltzmann problem, focusing on the case of muon tracks. We clarify our approach and summarize our results in a master formula. In \cref{sec:TransportEquation}, we solve the Boltzmann equation at second order in the small-energy-loss expansion (and at zeroth order in the small hard-scattering-rate expansion). In \cref{sec:QEDlosses}, we discuss the theoretical inputs entering the collisional integral, namely the cross-sections of the different radiative processes responsible for high-energy muon energy losses. In \cref{sec:MuonEnergyLoss:Fits}, we discuss the MC simulation of muon energy losses by which we extract the distributions of the parameters entering the resulting equation. In \cref{sec:FitToData}, we employ our formula to perform fits to \ic{} and \kmn{} data. Finally, in \cref{sec:conclusions}, we discuss our conclusions. The paper is augmented with two appendices: in \cref{app:qed-collisional} we derive from first principles the soft expansion of the collisional integral, review the necessary steps to solve the Fokker--Planck equation with constant coefficient and generalize the solution to weakly varying coefficient. In \cref{app:HardScatterings} we discuss the inclusion of the effect of hard scattering as small perturbations with respect to the soft propagation. 

\section{Setting up the Boltzmann problem}
\label{sec:setup}

In this section we set up the transport equations for muon tracks. As argued above, these constitute a large fraction of the total event rate and require a careful treatment of the muon energy loss. The discussion here can be straightforwardly adapted to other event topologies.

In the language of Boltzmann equations, the expected number of muons at the detector is expressed in terms of the muon distribution function, $f_\mu$, as defined in \cref{eq:N_from_fmu}, that is, normalized according to standard phase-space conventions. 
The muon distribution satisfies the equation
\beq\label{eq:full-boltzmann}
\frac{\dd f_\mu(t,\vec x,\vec p)}{\dd t}
\equiv
\deriv{f_\mu}{t}
+
\vec v_\mu\cdot\nabla f_\mu
=
C_{\rm weak}\!\left[t,\vec x,\vec p;\phi_\nu^\oplus\right]
+
C_{\rm QED}\!\left[t,\vec x,\vec p;f_\mu\right]\,,
\eeq
where on the left-hand side we have assumed free propagation between collisions, $\dot{\vec x}=\vec v_\mu$ and $\dot E=0$. In the ultra-relativistic regime relevant for this work, muons propagate with velocity $v_\mu\simeq c$, which we assume throughout the following. Since source, detector exposure and medium vary on timescales which are larger than $R_\oplus/c\approx 0.01\, {\rm sec}$, we seek stationary solutions and set $\partial_t f_\mu=0$ throughout this work. We set $c=1$ hereafter.

On the right-hand side, two terms drive the evolution of the muon distribution. On the one hand, the source term $C_{\rm weak}$ describes the production of muons by neutrino interactions in the Earth,
\beq\label{eq:C_weak}
C_{\rm weak}\!\left[t,\vec x,\vec p\,;\phi_\nu^\oplus\right]
=
\frac{(2\pi)^3}{E^2}
n_N(\vec x)
\int \dd E_\nu \dd\Omega_\nu\,
D_\nu(E_\nu,\vec x,\Omega)\,
\phi_\nu^\oplus(E_\nu,\Omega)
\left(
E^2\frac{\dd\sigma^{\rm{CC}}_{\nu N}}{\dd^3p}
\right)\,,
\eeq
where the flux of a particle species $a$ is defined as
\beq
(2\pi)^3\phi_a = E^2 v f_a\,.
\eeq
In \cref{eq:C_weak}, $n_N(\vec x)$ is the density of the material traversed by the neutrino, while
$D_\nu\equiv \phi_\nu/\phi_\nu^\oplus $ is the attenuation factor, defined as the ratio between the neutrino flux at the point $\vec x$ and the flux at the Earth surface. It depends on energy, direction, and traveled distance through the corresponding column density $L\lp\vec{x}; \Omega\rp$. Schematically,
\beq\label{eq:atten}
D_\nu \approx \exp\lp- n_N \sigma_{\nu N}\lp E_\nu\rp L\lp\vec{x}; \Omega\rp\rp\,.
\eeq
The CC cross-section of neutrino DIS onto nucleons is controlled by parton distribution functions (PDFs) at $Q^2\sim M_W^2$ and very small Bjorken-$x$, where extrapolations below the HERA region introduce the dominant uncertainty~\cite{Cooper-Sarkar:2011jtt,Cooper-Sarkar:2011jhk,Gauld:2015kvh}. In the following we use the {\tt MadGraph v3.5.1}~\cite{Alwall:2014hca} simulated cross section below 10 PeV, while above this energy we parametrize it as  
\begin{equation}\label{eq:crossection-Above10PeV}
\sigma_{\nu N}^{\rm CC}(E_\nu)= \sigma_0 \lp \frac{E_\nu}{E_0}\rp^\lambda = 1.48\times10^{-33}~{\rm cm}^2 \times \lp \frac{E_\nu}{10~{\rm PeV}} \rp^\lambda\,.
\end{equation}
As explained in Ref.~\cite{Palmisano:2025abd}, this parametrization fixes the $\sigma_0$ by matching the cross section to {\tt MadGraph} result at $E_\nu=E_0=10\,\rm{PeV}$ with the default LHA-PDF~\cite{Buckley:2014ana}. At higher energies we allow $\lambda$ to vary with the {\tt MadGraph} result closely following the $\lambda=0.4$ case. 

On the other hand, the QED collisional term, $C_{\rm QED}$, accounts for the energy degradation during propagation, thus it depends on the muon distribution itself. In full generality, $C_{\rm QED}$ is an integral operator in momentum space. In the high-energy limit, however, angular deflections can be neglected and the muon direction is approximately fixed by the parent neutrino direction. The collisional integral then becomes effectively one-dimensional in energy. The derivation and regime of validity of this approximation is discussed in \cref{app:qed-collisional}. 

Let us define the fractional energy loss $y\equiv \Delta E/E$. Keeping only the explicit dependence on the muon energy, the QED collisional operator can be written as
\beq\label{eq:C_QED_y_general}
C_{\rm QED}[E,\hat p;\phi_\mu]
=-\int_0^1 \dd y\,
\frac{\dd\Gamma(E,y)}{\dd y}\,
\phi_\mu(E,\hat p)+
\int_0^1 
\frac{\dd y}{1-y}
\frac{\dd\Gamma\left(E_y,y
\right)}{\dd y}
\phi_\mu
\left(E_y,\hat p
\right) .
\eeq
where we defined $E_y=E/(1-y)$, while ${\dd\Gamma(E,y)}/{\dd y}= n_T\, \dd\sigma(E,y)/\dd y$ is the differential rate in the ultra-relativistic limit. Depending on the process, the integral spans from $0$ to $1$ or from an IR threshold ${y_{\rm min}}$ to 1. The first term in \cref{eq:C_QED_y_general} is the loss term, removing muons from energy $E$. The second term is the gain term: a muon observed with energy $E$ can originate from a higher-energy muon with initial energy $E_y$, which lost a fraction $y$ of its energy. Therefore the differential rate in the gain term should in general be evaluated at the incoming energy.  

In order to reduce the full integro-differential collision equation to a local differential equation, we exploit the fact that a large fraction of QED scatterings are soft, namely characterized by fractional energy losses
\(y\ll1\).  In this regime the shifted energy in the gain term can be expanded around the observed energy, leading to a Kramers--Moyal expansion of the collisional operator~\cite{Risken}. This approximation is robust, as long as the scatterings are dominated by small
inelasticities and catastrophic losses are rare.  At higher energies, bremsstrahlung and photonuclear interactions develop harder tails in \(y\) as we will discuss in \cref{sec:QEDlosses}, and
the accuracy of a purely local expansion must be assessed with care.

To make this separation explicit, one may introduce a cutoff \(y_{\rm cut}\) and
write 
\begin{equation}\label{eq:hardvssoft}
C_{\rm QED}
=
C_{\rm QED}^{\rm soft}
+
C_{\rm QED}^{\rm hard}\,,    
\end{equation}
where the soft operator receives contributions from \(y<y_{\rm cut}\), while the
hard operator contains the complementary region \(y_{\rm cut}<y<1\).  The soft
piece can be expanded in derivatives of the energy distribution, giving the
Fokker--Planck operator used below.  In the main analysis we take
\(y_{\rm cut}\to1\), thereby using the local expansion as an approximation to
the full QED collision term.  The dependence on \(y_{\rm cut}\), and the
correction induced by the non-local hard operator, are discussed in \cref{app:HardScatterings}.

At leading order, electromagnetic losses enter only through the first moment of the transfer distribution, resulting in a deterministic evolution of the average muon energy. The soft expansion can be consistently extended to second order. The resulting correction is controlled by the second moment of the energy-transfer distribution and accounts for the stochastic nature of muon energy losses, inducing a diffusion term in energy space, driving the evolution of the variance of the muon energy distribution. 
Leaving the details of such expansion in \cref{app:qed-collisional}, upon defining
\beq\label{eq:b-d-def}
b_\mu(E) = \int\limits_{y_{\rm min}}^1 ~\dd y~y~\frac{\dd\Gamma(E,y)}{\dd y}\,, \qquad d_\mu(E) = \int\limits_{y_{\rm min}}^1 ~\dd y~y^2~\frac{\dd\Gamma(E,y)}{\dd y}\,,
\eeq
the stationary Boltzmann equation reduces to 
\beq\label{eq:fokker-planck}
\deriv{\phi_\mu}{x}
-
\deriv{}{E}
\Big(
b_\mu(E) E\phi_\mu(E,x)
\Big)
-
\frac12
\frac{\partial^2}{\partial E^2}
\Big(
d_\mu(E) E^2\phi_\mu(E,x)
\Big)
=
\frac{E^2}{(2\pi)^3}C_{\rm weak}\,,
\eeq
that is, we have a drift-diffusion equation for the muon flux, where the coefficient $b_\mu(E)$ controls the drift of energy, while $d_\mu(E)$ controls its diffusion. Neglecting for simplicity the slow, logarithmic dependence of the coefficients on energy, \cref{eq:fokker-planck} has a well known and simple solution, which we derive in \cref{sec:TransportEquation}. The more involved solution taking into account the energy dependence is derived in \cref{sec:energydep}.  

It is worth noticing that Pawula's theorem~\cite{Pawula:1967zz} implies that a finite Kramers-Moyal truncation beyond second order cannot in general be interpreted as a positivity-preserving evolution equation for $f_\mu$. 
Indeed, the presence of \emph{hard} energy losses, with \(y\sim 1\), can in principle spoil the convergence of the soft expansion. In the extreme limit in which hard scatterings dominate the collision integral, the expansion in powers of \(y\) breaks down and one must retain the full integro-differential form of \cref{eq:C_QED_y_general}. However, since hard scatterings occur at a sufficiently small rate for muon energy losses in matter, their effect can still be incorporated perturbatively around the soft solution, provided the correction remains parametrically smaller than the leading positive distribution. We will follow this route in \cref{app:HardScatterings}.

\subsection{The differential event rate}
\label{sec:detector}
Once the Boltzmann problem is set and the muon distribution function can be computed, we must define from it an event rate that can be compared with experimental measurements. In order to do so, we need to model the detector. Let us assume a spherical detector with unit efficiency, which is perfectly absorbent, in such a way that each muon reaching it, or being produced inside it, is counted once and is then removed from the distribution.
This model can be implemented by adding to the Boltzmann equation a detector collisional term
\begin{equation}
C_{\rm det}[f_\mu]
=
-\Gamma_{\rm det}({\bf x},{\bf p})\,
f_\mu({\bf x},{\bf p},t)\,,
\end{equation}
where $\Gamma_{\rm det}$ only has support inside the instrumented detector volume, $V_{\rm det}=4\pi R_{\rm{det}}^3/3$. In principle, this function can be generalized to account for detector efficiency and geometric acceptance. We leave the implementation of a realistic detector response to future work, where it could be incorporated through dedicated detector simulations and experimental inputs. The perfectly absorbent limit that we take corresponds to $\Gamma_{\rm det}({\bf x},{\bf p})\to \infty$. The detected event rate is then identified with the total rate at which muons are removed by the detector,
\begin{equation}
\label{eq:dndt_sink}
\left.\frac{\mathrm d N_{\rm{tracks}}}{\mathrm d t}\right|_{\rm det}
=
\int_{V_{\rm det}}\mathrm d^3x\,\mathrm d^3p\,
\Gamma_{\rm det}({\bf x},{\bf p})\,
f_\mu({\bf x},{\bf p},t)\,.
\end{equation}
Integrating the stationary Boltzmann equation over the detector volume and over momenta, we obtain\footnote{Note that in the perfectly absorbent limit the detector cannot accumulate particles, so that $\frac{\mathrm d}{\mathrm dt}\int_{V_{\rm det}}\mathrm d^3x\,\mathrm d^3p\,f_\mu = 0$. As a consequence, the explicit time-derivative term does not contribute to the event-rate formula. This does not imply $\partial_t f_\mu=0$ in the Boltzmann equation: a time dependence may still arise from the source, detector response, or propagation medium. Such effects become relevant only if they vary on timescales comparable to the muon Earth-crossing time, $\tau_\oplus\sim R_\oplus/c\simeq2\times10^{-2}\,\mathrm{s}$.}

\begin{equation}
\label{eq:dndt_abs}
\left.\frac{\mathrm d N_{\rm{tracks}}}{\mathrm d t}\right|_{\rm det}
=
{\color{blueviolet} \int_{V_{\rm det}}\mathrm d^3x\,\mathrm d^3p\,
C_{\rm weak}[f_\mu]}
+
{\color{coral}\int_{\partial V_{\rm det}}\mathrm d^2\Sigma\,\mathrm d^3p\,
f_\mu\, \left(-{\bf v}\cdot\hat{\bf n}\right)}\,,
\end{equation}
with both terms being positive by virtue of the fact that for muons reaching the detector $\bf v \cdot \hat n < 0$. This equation is a precise version of \cref{eq:Nexpected:Sketch}: the first term counts muons produced directly inside the detector, while the second term features the soft detector volume, counting muons produced outside the detector and transported up to its boundary.

\begin{figure}
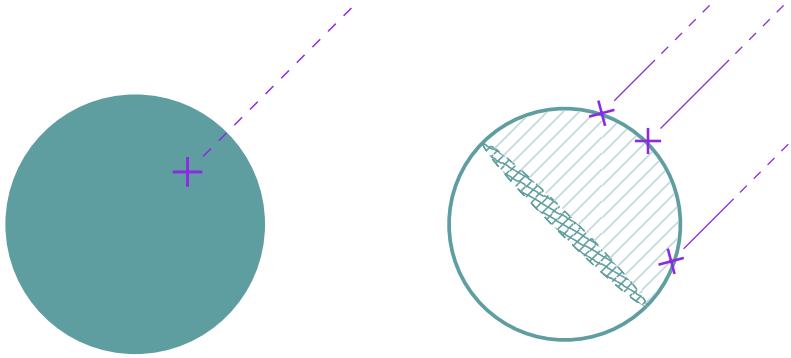

    \centering
    \begin{tabular}{cc}
         \inside{.35\linewidth}& \outside{.35\linewidth} \\ 
    \end{tabular}
    \caption{Muons produced inside (left) or reaching (right) the detector, are immediately detected with unit efficiency and removed. The detection rate of the former is thus proportional to the detector volume, whereas the detection rate of the latter si proportional to the projected area of the detector (hatch-filled equatorial circle).}\label{fig:sticky-detector}
\end{figure}

If the neutrino flux and attenuation vary negligibly across the detector scales, the inside contribution can be easily rewritten in terms of the impinging neutrino flux. For the outside contribution, we assume that muons arriving from a given direction are parallel to one another, and that the muon distribution varies slowly along the direction transverse to that of muon propagation. Then the surface integral reduces to the projected area of the detector along the direction of propagation (see \cref{fig:sticky-detector}). For a spherical detector of radius $R_{\rm det}$, $A_{\rm proj}=\pi R_{\rm det}^2$.

Thus the total differential event rate can be written as
\begin{equation}
\label{eq:dndt_final}
\frac{\mathrm d N_{\rm{tracks}}}
{\mathrm d t\,\mathrm d E\,\mathrm d\Omega}
\approx
{\color{blueviolet} V_{\rm det}\,
n_N^{\rm det}
\int \mathrm dE_\nu\,
\phi_\nu(E_\nu,\Omega)\vert_{{\rm det}}\,
\frac{\mathrm d\sigma_{\nu N}^{\rm CC}}
{\mathrm dE}}
+
{\color{coral} A_{\rm proj}\,
\phi_\mu(E,\Omega)\vert_{{\rm det}}}\,,
\end{equation}
where the muon flux at the detector boundary, $\phi_\mu(E,\Omega)\vert_{\rm det}$, is obtained by solving the transport equation and depends on the transport coefficients determined by the QED energy-loss rates and the Earth column density, as discussed in the following sections. \Cref{eq:rate:Complete} therefore provides a concrete framework to compute the contribution of soft-volume events. Throughout the paper we are going to identify the soft volume as a function of the muon energy as
\begin{equation}
{\color{coral} V_{\rm{soft}}(E)\left[ n_N \sigma_{\nu N}(E) \phi_\nu^\oplus(E)\right]= A_{\rm proj}\,
\phi_\mu(E,\Omega)\vert_{{\rm det}}}\,.
\end{equation}
Since $\phi_\mu(E,\Omega)$ in turn depends on the impinging neutrino flux, the soft volume will depend in general on the flux parameters.

We stress that the assumption of a perfectly absorbent detector generally overestimates the event rate, since it neglects finite detector efficiency, geometric acceptance, and event-selection cuts. In the simplest approximation, these effects can be parametrized by an overall efficiency factor $\epsilon_i$, which depends on the detector and on the specific event selection. This amounts to the rescaling $V_{\rm det}\to \epsilon_i V_{\rm det}$ and $A_{\rm proj}\to \epsilon_i A_{\rm proj}$. For instance, in \cref{sec:FitToData} we will derive the efficiency $\epsilon_{\rm IC\text{-}TG}$ relevant for the IceCube through-going muon analysis~\cite{Abbasi:2021qfz} by comparing our prediction with the official IceCube result. We will also show how the extracted efficiency matches the reduction of the number of events expected by the angular cut imposed on through-going tracks at IceCube. 

More generally, detector effects can be incorporated through energy- and direction-dependent response functions in \cref{eq:dndt_sink}. This would make the map from the impinging neutrino flux to the expected number of events fully consistent with the experimental analysis. Unfortunately, such response functions are not publicly available for most searches. We encourage the experimental collaborations to provide detector response maps for each analysis, so that their results can be fully reproduced and reinterpreted.

As a final remark, it is worth noticing that cascade events are easily captured by \cref{eq:dndt_final} by taking only the first term of the sum and summing CC contributions with electrons in the final state as well as NC events.

\subsection{Solving the transport equation}
\label{sec:TransportEquation}

In this section, we provide the solution to the drift-diffusion problem in \cref{eq:fokker-planck}.
In the language of Green's functions we can write the muon flux as
\beq\label{eq:solution-from-greens}
\phi_\mu(E,x) = \int_0^\infty \dd\xi \int_0^\infty \dd\varepsilon \, G(E, x; \varepsilon, \xi) \,\frac{\varepsilon^2}{(2\pi)^3}\, C_{\rm weak}\left[\varepsilon,\xi;\phi_\nu^\oplus\right]\,,
\eeq
which is convolving all the points and energies sourced by the weak current collision term with the QED propagation encoded in the Green's function $G(E, x; \varepsilon, \xi)$. The latter solves \cref{eq:fokker-planck}, sourced by an impulsive source $\delta(x-\xi)\delta(E-\varepsilon)$, and it is readily found to be
\be\label{eq:Greens}
G(E,x;\varepsilon, \xi) =  \frac{\theta\left(x-\xi\right)\theta\left(\varepsilon-E\right)}{\sqrt{2\pi E^2 d_\mu \left(x-\xi\right)}} \exp\left[-\frac{\left(\log\frac{\varepsilon}{E} - \left(b_\mu + \frac {d_\mu}2\right) \left(x-\xi\right)\right)^2}{2 d_\mu \left(x-\xi\right)}\right] \,.
\ee
Here we are neglecting the slow dependence on energy of $b_\mu$ and $d_\mu$, deferring the full solution to \cref{sec:energydep}. \cref{eq:Greens} is to be intended as the solution for fixed muon impinging angle $\Omega_p$, with $x$ being the corresponding column depth. The dependence of the neutrino flux at the crust on the angle in \cref{eq:dndt_final} will then determine the rate. In this work we focus only on diffuse fluxes, for which $\phi_\nu^\oplus$ is independent of the angle. For a point source, $\phi_\nu^\oplus \propto \delta^{(2)}\left(\Omega_\nu - \Omega_{\rm PS}(t)\right)$, in which case to compute the total number of expected events one must average the neutrino attenuation -- the only quantity with an explicit dependence on the angle in \cref{eq:C_weak} -- over the relevant time range \cite{Palmisano:2025abd}.

\Cref{eq:Greens} describes the propagation of a charged lepton in a medium from a point $\xi$ where it is produced with energy $\varepsilon$, to a point $x$ where it arrives with energy $E$. The transport takes place through the radiative processes described in \cref{sec:QEDlosses}, which are effectively encoded in the coefficients $b_\mu$ and $d_\mu$.\footnote{We are assuming here that the propagation happens in a homogeneous medium with constant density. Going beyond this assumption to consider inhomogeneous media or surface effects would require a more convolute expression. However, this is a reasonable assumption, given the typical experimental setups of neutrino telescopes.} When properly normalized, the function $G$ can also be interpreted as the conditional probability to have a muon with energy $E$ at location $x$ starting from a muon with energy $\varepsilon$ at location $\xi$.
This point of view will be taken in \cref{sec:MuonEnergyLoss:Fits} to infer the coefficients $b_\mu$ and $d_\mu$ from MC simulations. Once the Green's function is known, we can obtain the full solution to \cref{eq:fokker-planck} by convoluting $G$ with the source term, as in \cref{eq:solution-from-greens}.

Before proceeding, let us note that in the limit $d_\mu\to0$, the Green's function in \cref{eq:Greens} reduces to
\beq\label{eq:Greens:drift}
G(E,x;\varepsilon,\xi)
\xrightarrow{~d_\mu\to 0~}
\frac{\theta(x-\xi)}{E}
\delta\lp
\log\frac{E}{\varepsilon}
+
b_\mu(x-\xi)
\rp
=
\frac{\varepsilon}{E}
\delta\lp
\varepsilon-E(x)
\rp
\theta(x-\xi)\,,
\eeq
where we defined $E(x)=E\exp[b_\mu(x-\xi)]$. In the absence of diffusion, the lepton therefore propagates deterministically, losing energy according to the continuous slowing-down approximation. At energies below roughly $300\,{\rm GeV}$, where ionization losses dominate and are accurately described by the Bethe--Bloch equation~\cite{Groom:2001kq}, the transport is indeed well approximated by the drift term alone. Replacing \cref{eq:Greens:drift} into \cref{eq:solution-from-greens} reproduces the solution derived in Ref.~\cite{Palmisano:2025abd}. In the following we shall refer to this approximation as the \emph{drift} limit.

We can now rewrite the $C_{\rm weak}$ term from \cref{eq:C_weak} in a more convenient form. The lepton is produced by a neutrino of the same flavor in a charged current (CC) deep inelastic scattering (DIS), with energy $\varepsilon = (1-\yw)E_\nu$. Here $\yw$, usually called inelasticity in the context of DIS, represents again the energy loss of the process, although for CC the neutrino disappears, converting to a lepton; the subscript on $\yw$ is included to avoid confusion with the analogous quantity in QED processes. Furthermore, at UHE we can safely assume the lepton to be collinear with the neutrino, as the transverse deviation of the lepton is inversely proportional to its boost, $\Delta_\theta \propto m_\ell/\varepsilon\ll1$. We can then write 
\begin{equation}\label{eq:p_of_y}
 \frac{1}{\sigma_{\nu N}} \frac{{\rm d}\sigma_{\nu N}}{{\rm d}^3p} = \delta^{(2)}(\Omega_\nu - \Omega_p) \frac{(1 - \yw)}{\varepsilon^3} \py\,,
\end{equation}
where we have defined the inelasticity distribution as $\py\equiv\sigma^{-1}\dd\sigma/\dd \yw$. While in general this distribution needs to be computed numerically, e.g. using \texttt{MadGraph} simulations, in the UHE regime it can be computed analytically, finding that it is almost independent on the neutrino energy (see Ref.~\cite{Palmisano:2025abd} for the derivation). Replacing \cref{eq:p_of_y} in \cref{eq:C_weak} we can rewrite the weak collisional term as
\beq\label{eq:C_weak:Simplified}
\frac{\varepsilon^2}{\lp2\pi\rp^3} C_{\rm weak} = n_N(\xi) \int_0^1 \frac{\dd \yw \, \py}{1-\yw} D_\nu\left(\varepsilon_{\yw},\xi,\Omega\right) \sigma_{\nu N}^{\rm{CC}}\left(\varepsilon_{\yw}\right) \phi_\nu^\oplus\left(\varepsilon_{\yw}\right)\,,
\eeq
where $\varepsilon_{\yw}=\frac\varepsilon{1-\yw}$.
The dependence of the target density from the position $n_N(\vec{x})$ tracks the Earth's profile density. We employ the preliminary Earth model (PREM)~\cite{Dziewonski:1981xy}, in which the Earth is modeled as a series of concentric shell of constant density; in this model, the neutrino and lepton propagation in the medium depends only on its incoming polar angle with respect to the experiment horizon. An in-depth quantification of these terms has been carried out in Ref.~\cite{Palmisano:2025abd}, with particular attention to the anomalous event at \kmn.

Finally, the rate of muon events at detector can be written in an explicit form, under the same assumptions leading to \cref{eq:dndt_final}, as
\be\label{eq:rate:Complete}
\begin{split}
\left\langle \frac{\dd N}{\dd t \dd E \dd\Omega}\right\rangle& =
{\color{blueviolet} V_{\rm det} 
 \frac{E^2}{(2\pi)^3} C_{\rm weak}\left[
 E; \phi_\nu^\oplus\right] }+ \\
&\hspace{1.cm}+ {\color{coral}  A_{\rm proj}\int_{0}^{x}\dd\xi\int_E^\infty \dd\varepsilon \,G(E,x;\varepsilon,\xi) \frac{\varepsilon^2}{(2 \pi)^3} C_{\rm weak}\left[\xi, \varepsilon; \phi_\nu^\oplus\right]}\,.
\end{split}
\ee
In the second integral, the Green's function is evaluated on the detector surface, $x = L(\Omega_\nu) - R_{\rm det}$, where we let $L(\Omega_\nu)$ be the column depth of the center of the detector\footnote{Neglecting the size of the detector compared to the column depth of the detector, an approximate expression for the latter is \[  L(\Omega_p) = (R_\oplus - h_{\text{LAB}}) \big(\sqrt{\sin^2\delta -(1-R_\oplus^2/(R_\oplus-h_{\rm LAB})^2)} -\sin\delta\big)\,,\] where $\delta\in\left[-\frac{\pi}{2},\frac{\pi}{2}\right]$ is the elevation from the horizon, $R_\oplus$ is the Earth radius, and $h_{\text{LAB}}$ is the shortest distance between the crust and the center of the detector.} 
and $R_{\rm det}$ be its radius, under the assumption that the detector is a sphere and that the muon flux does not depend on the coordinate transverse to that of propagation. The integral limits are the Earth's surface at $\xi=0$ and the detector surface at  $\xi = x = L(\Omega_\nu) - R_{\rm det}$.

The multidimensional integral in \cref{eq:rate:Complete} does not admit a simple analytical evaluation in full generality and must therefore be treated numerically. Nevertheless, the resulting computation is sufficiently inexpensive that event rates in multiple energy and angular bins can be obtained within seconds on standard hardware. A public implementation of the framework presented in this work is currently in preparation and will be described in a dedicated software release~\cite{Palmisano:xxx}.

\subsection{Estimation of the soft volume}\label{sec:soft-volume}

Here we simplify the integral in \cref{eq:rate:Complete} by introducing a set of controlled approximations that make the physical origin of the soft volume transparent and allow its scaling relative to the detector volume to be estimated analytically.

First, we retain the detector model introduced in \cref{sec:detector}, namely a spherical and perfectly absorbent detector. Second, we parameterize the ultra-high-energy charged-current neutrino cross section as $\sigma_{\rm CC}(E_\nu)=\sigma_0(E_\nu/E_{\nu,0})^\lambda$, where $\lambda\simeq0.4$ \cite{Cooper-Sarkar:2011jtt}. This scaling follows from the fact that UHE neutrinos probe the nucleon PDFs at very small Bjorken-$x$, where the dominant parton distributions approximately behave as $x^{-\lambda}$. Under this assumption the inelasticity distribution of neutrino CC interactions, $\py$, can be computed analytically, as discussed around \cref{eq:p_of_y} and in Ref.~\cite{Palmisano:2025abd}. In practice, since $\langle \yw\rangle\simeq0.2$, the interaction is already close to the elastic limit, and the effect of replacing the full distribution by its average value is only at the percent level. Third, although the formalism applies to arbitrary neutrino fluxes, for comparison with the IceCube diffuse-flux measurements we assume a single power-law spectrum, $\phi_\nu\propto(E_\nu/100,{\rm TeV})^{-\gamma}$.

Finally, we approximate the neutrino attenuation with a step function: neutrinos arriving from above the horizon are taken to propagate freely, whereas those arriving from below are assumed to be completely absorbed. 
In this limit only the matter density in the vicinity of the detector, denoted by $n_N^{\rm near}$, affects charged-lepton propagation. Since this region is typically composed of a single material (ice for IceCube and seawater for KM3NeT), we further approximate $n_N^{\rm near}$ as constant.

Within these assumptions, the weak collisional integral reduces to 
\beq
C_{\rm weak} = n_N \sigma_{\nu N}(E) \phi_\nu^\oplus(E) \lp\frac\varepsilon E\rp^{\lambda-\gamma}\,,
\eeq
and the integral in \cref{eq:rate:Complete} can be rewritten as 
\begin{equation}\label{eq:generalsoft}
V_{\rm{soft}}(E)=A_{\rm{proj}}\int_0^\infty d\xi\int_E^\infty d\varepsilon \, G(E,L;\varepsilon,L-\xi)\left(\frac{\varepsilon}{E}\right)^{\lambda-\gamma}    
\end{equation}
where the integral over the physical distance can be pushed to infinity as long as the integrand vanishes sufficiently fast, making the depth $L$ at which the Green's function is evaluated immaterial. Focusing on the drift limit, the Green's function is the one in \cref{eq:Greens:drift} and we find
\beq\label{eq:AnalyticSolution:Drift}
V_{\rm{soft}}(E)\vert_{\rm{drift}}=A_{\rm{proj}}\int_0^\infty d\xi e^{-b_\mu A\xi} = \frac{\pi R_{\rm det}^2}{b_\mu}\frac{1}{A}\,,
\eeq
with $A=\gamma-\lambda-1>0$. The formula above shows that the soft volume is enhanced not just by the muon range, $1/b_\mu$, but also by the spectral penalty for producing a higher-energy parent neutrino, corrected by the Jacobian of the drift Green function.
From this formula one can read an estimate of the soft volume discussed in \cref{eq:Nexpected:Sketch} and \cref{eq:dndt_abs}. 
The total rate of events per energy bin receives contributions from inside the detector proportionally to its volume, as would be expected. The transport of muons, however, enhances this volume by roughly its surface area times the typical distance by which a muon of energy $E_\mu$ propagates before losing enough energy in radiative processes, namely $1/b_\mu$. 

As a figure of merit, we can estimate
\beq
\frac{V_{\rm{tot}}}{V_{\rm{det}}}\approx\left(1+\frac{3}{4 R_{\rm{det}} b_\mu}\right)\,
\eeq
for the IceCube measured diffuse flux, which has $A\approx1$ (see \cref{sec:FitToData}). Considering the instrumented volume of \ic{}, $V_{\rm det}^{\rm IC} \approx 1 \,{\rm km}^3$, corresponding to a sphere of radius $R_{\rm det}^{\rm IC} \approx  0.62\,\rm km$, and taking $b_\mu\approx 0.35/{\rm km}$  for a muon of $1\,\PeV$ (see \cref{tab:b-d}), the \emph{soft} volume is about four times as large as the instrumented volume. For a smaller instrumented volume, such as \kmn{}, still under construction, the ratio is even larger, reaching about $7.5$ with the same approximations.

In the diffusion case, the soft volume contribution can be obtained by replacing the Green's function from \cref{eq:Greens} in \cref{eq:generalsoft}, and can be written as
\beq\label{eq:AnalyticSolution:Diffusion}
V_{\rm{soft}}\vert_{\rm{diff}}(E)= \frac{A_{\rm{proj}}}{2} \int\limits  _0^\infty\dd\xi e^{-b_\mu A'\xi}\left(1+\erf(\sqrt{B\xi})\right)=
\frac{\pi R_{\rm det}^2}{2b_\mu A'}
\left(
1+\sqrt\frac{b_\mu}{{B+A' b_\mu}}
\right)\,.
\eeq
Here we defined $A'= A(1-\frac{d_\mu}{2b_\mu}(A-1))>0$ and $A'/B^2 = 2 A d_\mu/b^2_\mu$, up to subleading corrections in $d_\mu/b_\mu$. In the soft limit, $d_\mu/b_\mu\ll1$ and the correction to the soft volume in the diffusion case can be written as
$V_{\rm{soft}}\vert_{\rm{diff}}\simeq V_{\rm{soft}}\vert_{\rm{drift}}\left(1-\frac{d_\mu}{2b_\mu}\right)$,
showing that diffusion gives a negative correction to the soft volume proportional to $d_\mu/b_\mu$, which at leading order is independent on the details of the flux. This can be understood as the effect of stochastic broadening that redistributes events away from the deterministic drift trajectory at fixed observed muon energy.

\subsection{Tau-induced muon tracks}
\label{sec:TauInducedMuons}

We now estimate the contribution of tau leptons to the observed muon-track
sample. In the energy range of interest, tau propagation is mostly controlled
by decay rather than by electromagnetic energy losses. The tau decay length is
\beq
\ell_\tau(E_\tau)
\simeq
5~{\rm km}
\left(
\frac{E_\tau}{100~{\rm PeV}}
\right) ,
\eeq
while the QED range is parametrically larger, with
$b_\tau^{-1}\simeq 35~{\rm km}$ around the PeV scale. Therefore, for
$E_\tau\lesssim {\rm few}\times 100~{\rm PeV}$, we can neglect tau energy
losses and take $b_\tau\to0$.

The tau contribution to muon tracks is suppressed by the branching ratio ${\cal B}_{\tau\to\mu}\simeq0.17$ and by the fact that the daughter muon carries only a fraction of the tau energy. We approximate the decay kinematics by $E_\tau \simeq q E_\mu$ with
$q\simeq 3$. At fixed observed muon energy, the tau-induced contribution is therefore weighted by ${\cal B}_{\tau\to\mu} q^{-A}$. The factor $q^{-A}$ includes the energy dependence of the neutrino flux and cross section, together with the Jacobian from $E_\tau$ to $E_\mu$.

For both starting and through-going events we have $N_{\tau\to\mu}/
N_{\nu_\mu}\simeq
{\cal B}_{\tau\to\mu}q^{-A}\approx 0.057$ for a flavor-democratic flux given that $A\approx1$. We thus expect tau-induced muons to contribute at $5\%$ level to a given muon bin and hence we neglect them in what follows. Note however that the suppression strongly depends on the steepness of the flux, the $\g$ exponent, and this estimation can drastically change in exotic scenarios with peaked fluxes.

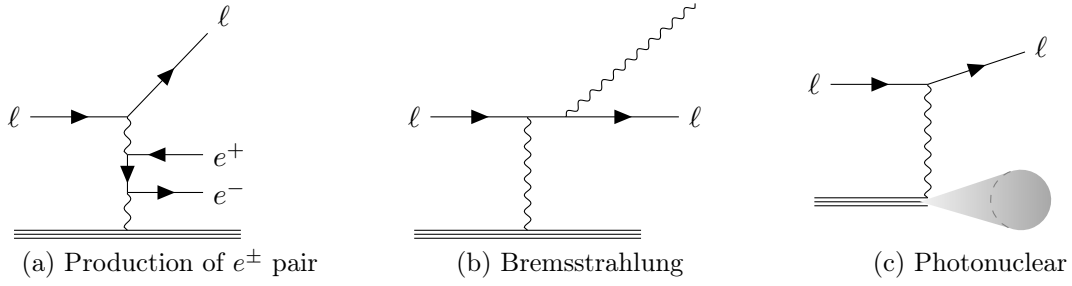
\begin{figure}[t]
\begin{subfigure}{.3\textwidth}
\begin{tikzpicture}
    \begin{feynman}
        \vertex (a) {\(\ell\)};
        \vertex [right=of a] (b) ;
        \vertex [below=.5 cm of b] (n1) ;
        \vertex [below= .5cm of n1] (n2) ;
        \vertex [below = .5 cm of n2] (c) ;
        \vertex [above right = of b] (f1) {\(\ell\)};
        \vertex [above = of f1] (g);
        \vertex [right = 1cm of n1] (e1) {\(e^+\)};
        \vertex [right = 1cm of n2] (e2) {\(e^-\)};
        \vertex [left = of c] (d1);
        \vertex [right = of c] (d2);
        \diagram* {
        (a) -- [fermion] (b) -- [fermion] (f1),
        (b) -- [boson] (n1) -- [fermion] (n2) -- [boson] (c),
        (n1) -- [anti fermion] (e1),
        (n2) -- [fermion] (e2),
        (d1) -- [ plain] (d2)
        };
    \end{feynman}
    \draw[transform canvas={yshift=-3pt}] (d1) -- (d2);
    \draw[transform canvas={yshift=-1.5pt}] (d1) -- (d2);
\end{tikzpicture}
\caption{Production of $e^{\pm}$ pair}
\end{subfigure}
\hfill
\begin{subfigure}{.3\textwidth}
\begin{tikzpicture}
    \begin{feynman}
        \vertex (a) {\(\ell\)};
        \vertex [right=of a] (b) ;
        \vertex [right=.5cm of b] (m);
        \vertex [below=of b] (c) ;
        \vertex [right = of m] (f1) {\(\ell\)};
        \vertex [above = of f1] (g);
        \vertex [left = of c] (d1);
        \vertex [right = of c] (d2);
        \diagram* {
        (a) -- [fermion] (b) -- [plain] (m) -- [fermion] (f1),
        (b) -- [boson] (c) [blob],
        (m) -- [boson] (g),
        (d1) -- [ plain] (d2)
        };
    \end{feynman}
    \draw[transform canvas={yshift=-3pt}] (d1) -- (d2);
    \draw[transform canvas={yshift=-1.5pt}] (d1) -- (d2);
\end{tikzpicture}
\caption{Bremsstrahlung}
\end{subfigure}
\hfill
\begin{subfigure}{.3\textwidth}
\begin{tikzpicture}
    \begin{feynman}
    \vertex (a) {\(\ell\)};
    \vertex [right=of a] (b);
    \vertex (f1) at ($(b) +  (1.5cm, .5cm)$) {\(\ell\)};
    \vertex [below =of b] (c);
    \vertex [left=of c] (n0) ;
    \vertex [right=of c] (n1);
    \vertex [above = .3cm of n1] (n2);
    \vertex [below=.4 cm of n1] (n3);
    \vertex [above=.2 cm of n2] (n4);

    \diagram* {
      (a) -- [fermion] (b) -- [fermion] (f1),
      (b) -- [boson] (c),
      (n0) -- [plain] (c)
    };
    \end{feynman}
    \draw[transform canvas={yshift=-3pt}] (n0) -- (c);
    \draw[transform canvas={yshift=-1.5pt}] (n0) -- (c);
  \coordinate (cv) at ($(c)+(-3pt,-1.5pt)$); 
  \coordinate (ctop) at ($(cv)+(1.2cm,0.4cm)$);
  \coordinate (cbot) at ($(cv)+(1.2cm,-0.4cm)$);

  
  \shade[left color=gray!10, right color=gray!70, middle color=gray!40]
    (cv) -- (ctop) arc[start angle=110, end angle=-110, radius=0.4cm] -- cycle;

  \draw[color=gray, dashed] (ctop) arc[start angle=110, end angle=250, radius=0.4cm];
\end{tikzpicture}
\caption{Photonuclear}
\end{subfigure}
\caption{Representative Feynman diagrams of the three radiative processes, relevant at high energy for the charged lepton energy loss in a medium.}
\label{fig:diagrams-radiative}
\end{figure}

\section{Transport coefficients from QED}
\label{sec:QEDlosses}
The QED collisional integral encodes the charged lepton energy losses in the propagation medium. 
Generally, as a charged lepton propagates through matter, its energy is degraded by electromagnetic interactions with the constituents of the medium. These processes are schematically of the form
\be\label{eq}
\ell(\vec p) + T \to \ell(\vec p^{\,\prime}) + X\,,
\ee
where $\ell$ denotes a charged lepton and the target $T$ corresponds either to the full nucleus in coherent scatterings or to individual nucleons in hard interactions. The inclusive final state $X$ may contain photons, lepton pairs, nuclear excitations and hadrons. 

\begin{figure}[t]
\centering
    \includegraphics[width=0.45\linewidth]{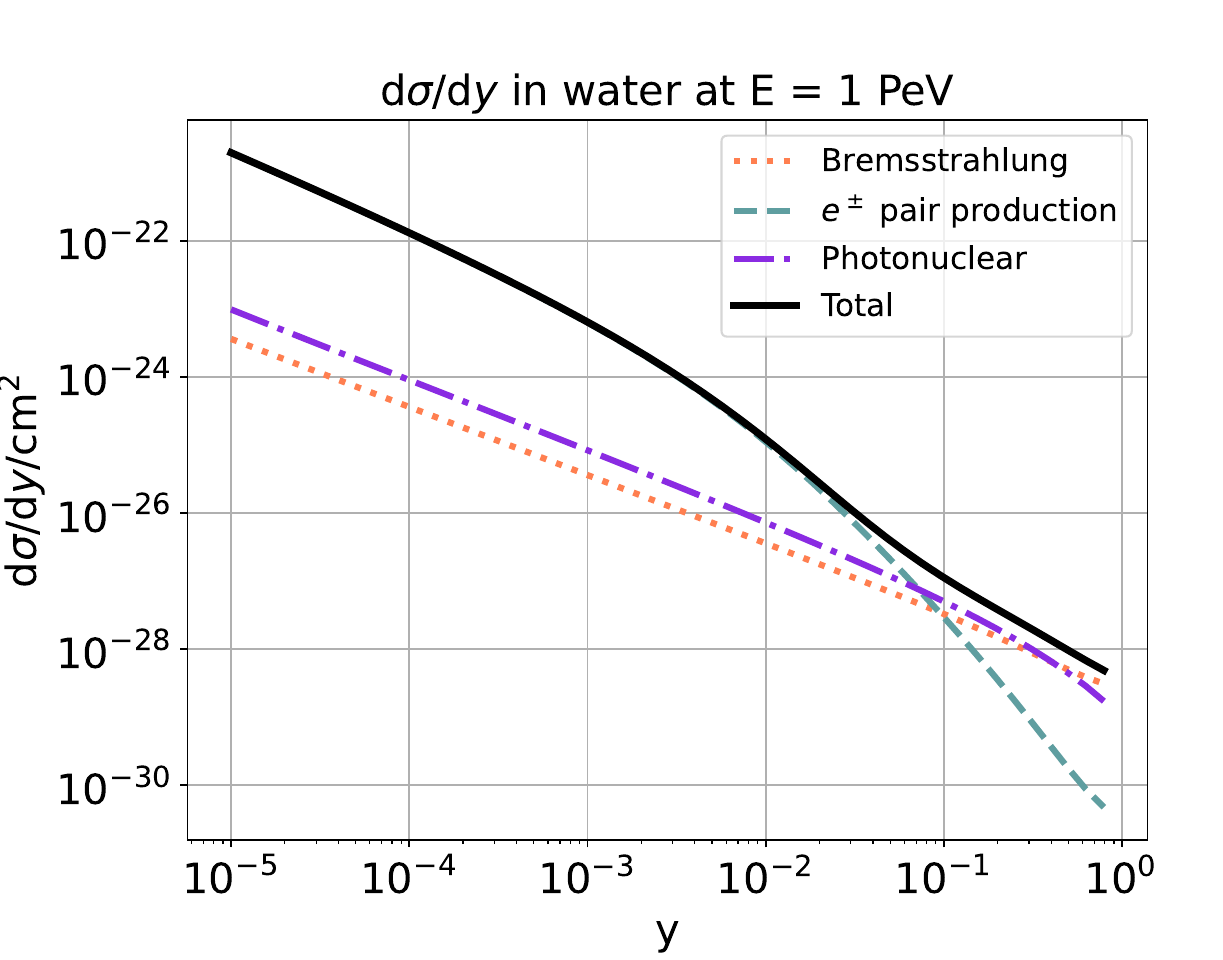} \\
    \includegraphics[width=0.45\linewidth]{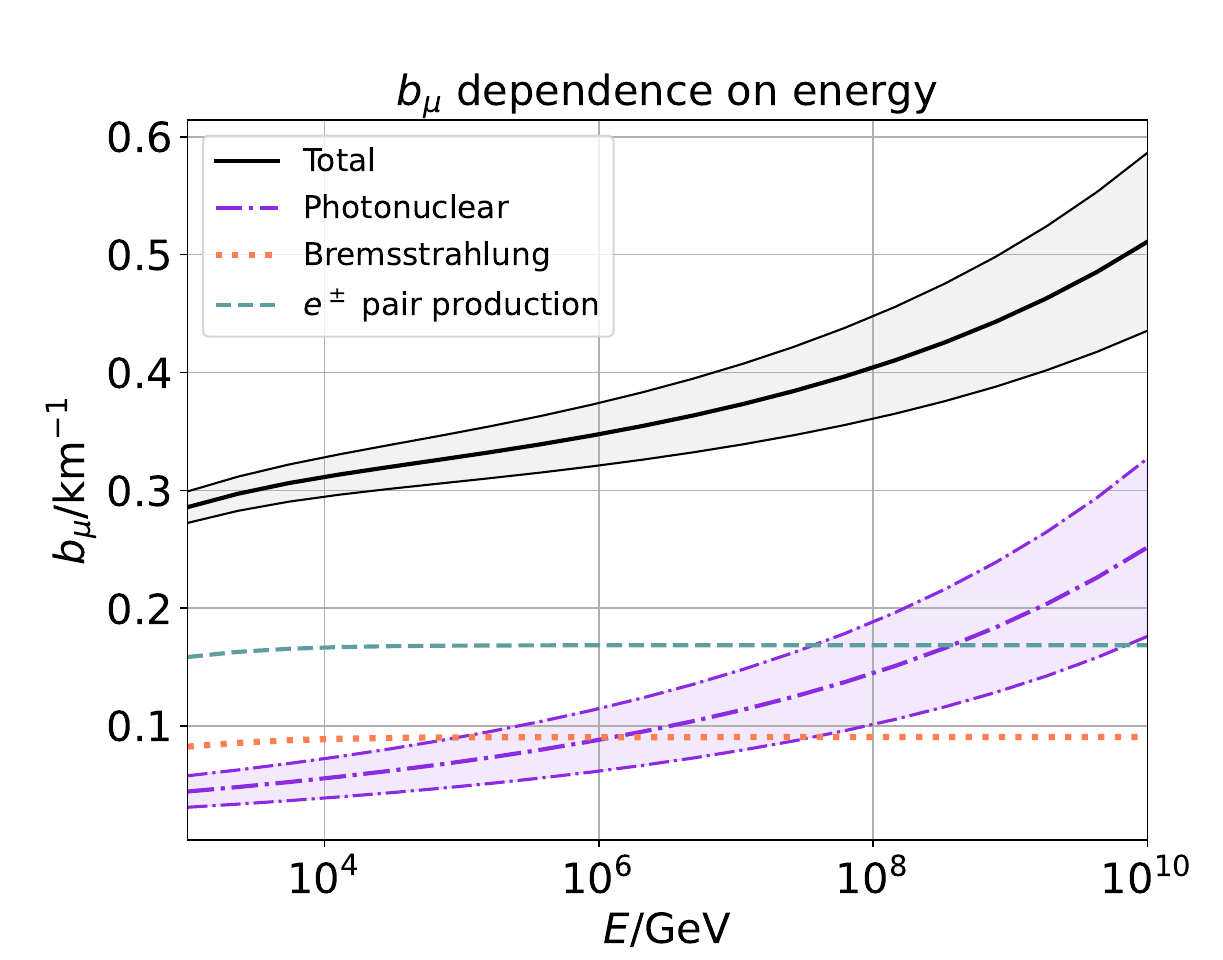}\hfill 
    \includegraphics[width=0.45\linewidth]{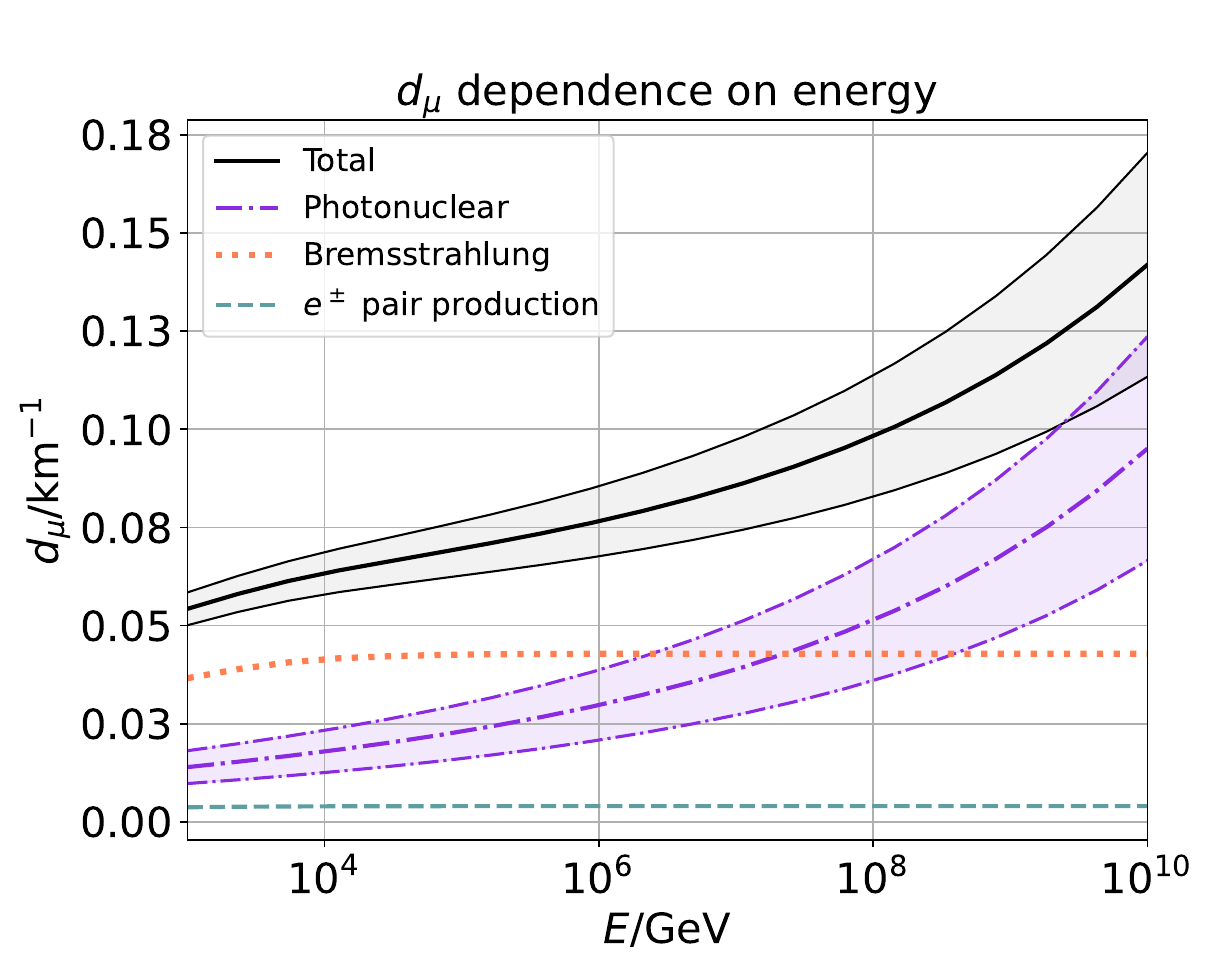}
    \caption{{\bf Top}: Energy loss distributions, $\dd\sigma/\dd y$, for a muon with energy $E_\mu=1$ PeV propagating in water. The dashed green, dotdashed purple and dotted red lines indicate the $e^\pm$ pair production, photonuclear and bremsstrahlung terms, respectively, while the solid black line shows the total. {\bf Bottom}: drift coefficient $b_\mu$ (left) and diffusion coefficient $d_\mu$ (right) as function of the muon energy, with same color code as top plot for the single terms. The purple band indicates the uncertainty on the photonuclear rate, see text for details, which reflects on the uncertainty on the total coefficients, shown by the black band.}
    \label{fig:dsdy}
\end{figure}

In the SM, at the energies of interest for this work, the energy losses are dominated by the three radiative processes shown in \cref{fig:diagrams-radiative}. To connect with the existing literature, we briefly summarize the main features of these processes, while in the numerical analysis of \cref{sec:MuonEnergyLoss:Fits} we employ the full cross-sections implemented in Ref.~\cite{Koehne:2013gpa} and references therein. In particular, we are interested in their energy-loss spectra, $\dd\sigma/\dd y$, whose shape determine the goodness of the expansion in \cref{eq:fokker-planck}. These spectra are shown in the left panel of \cref{fig:dsdy}, for $1\,\PeV$ muons in water.
In \cref{eq:tabellalosses} we also summarize the values of the relevant quantities discussed in this section, for muons of two reference energies, $E_\mu=1,100\,\PeV$, traversing water\footnote{For neutral compounds, the total QED rate and its moments are an incoherent sum of the single interaction on the constituent atoms. For water $b_{\rm th}^{\rm wtr} = 2b_{\rm th}^{\rm H} + b_{\rm th}^{\rm O}$, where $Z_H = A_H = 1$ and $Z_O = 8$,$A_O=16$ (and similarly for $d_{\mu}$).} (density $\rho_{\rm wtr}=1.02~\gcm$ ). 

For all three processes, the interaction of a high-energy charged particle with matter depends on the nuclear response, which is encoded in suitable nuclear form factors or structure functions. This introduces a theoretical uncertainty that becomes increasingly relevant at the very high energies considered here. Among the three processes, photonuclear interactions are the most affected by these uncertainties and therefore provide the dominant contribution to the theoretical error budget~\cite{Koehne:2013gpa}. In the following, we describe the three contributions in turn.
 
\paragraph{\bf Lepton pair production (PP):}
$\ell^\pm + T\to\ell^\pm +\ell^{\prime +}+\ell^{\prime -}+T$.

Pair production proceeds through a $t$-channel photon-fusion process in which the incoming lepton $\ell$ emits a pair of charged leptons $\ell'$ (left diagram in \cref{fig:diagrams-radiative}). The cross-section scales as $\alpha^4$, but is enhanced by the propagator of the produced lepton as $m_{\ell'}^{-2}$. As a consequence, $e^\pm$ pair production is always dominant, while heavier pair channels are strongly suppressed, e.g. $m_e^2/m_\mu^2\sim10^{-4}$; in the following, PP refers exclusively to $e^\pm$ pair production. For any lepton, the spectrum is kinematically constrained from below by the lepton mass, that is, we have $y_{\rm min} = 4m_\ell'/E_\ell$.
As shown in \cref{fig:dsdy}, PP is the dominant radiative process for muons at low exchanged momentum. And it can also be trustfully described by the soft expansion in \cref{eq:fokker-planck} since the energy loss spectrum vanishes at large $y$, behaving approximately as $(1-y)/y$. 
In \cref{sec:MuonEnergyLoss:Fits}, we use the Kelner-Kokoulin-Petrukhin (KKP) parametrization of the nuclear and atomic response~\cite{Koehne:2013gpa}, consistently with modern MC implementations. 

\paragraph{\bf Bremsstrahlung (B):}
$\ell^\pm + T\to\ell^\pm +\gamma+T$.

Bremsstrahlung is the emission of a photon in the EM field of the nucleus (center diagram in \cref{fig:diagrams-radiative}). The dominant contribution arises from coherent recoil off the full nucleus, while sub-leading inelastic contributions originate from atomic electrons or excited nuclear states. The cross-section scales approximately as $m_\ell^{-2}$, implying a hierarchy of several orders of magnitude between Bremsstrahlung energy losses for electrons, muons, and taus. As for PP, we use the KKP parametrization in our numerical evaluations \cite{Koehne:2013gpa}. Note that, since in principle there is no lower value $y_{\rm min}$ for the emitted photon energy, the cross-section is IR-divergent. In this sense, the only quantities that have physical meaning are the energy loss rates, $\int\dd y~y^n~\dd\sigma/\dd y$ for $n>1$.
The energy loss spectrum behaves roughly as $1/y$, meaning that catastrophic losses with $y\sim1$ are allowed, potentially breaking the soft expansion in \cref{eq:fokker-planck}. 

\paragraph{\bf Photonuclear scattering (PN):}
$\ell^\pm + T\to\ell^\pm +X$.

Photonuclear interactions (right diagram in \cref{fig:diagrams-radiative}) refer to the ensamble of inelastic EM interactions of the lepton with the nuclei. Unlike PP and B, where coherent scattering on the full nucleus dominates, photonuclear interactions are intrinsically incoherent. Depending on the momentum transfer, the final state $X$ may correspond to excited nuclear states at small $Q^2$, ejected nucleons for $Q^2\sim \Lambda_{\rm QCD}^2$, or hadronic jets at large $Q^2$. Since the cross section is dominated by small momentum transfers, where the nuclear response is intrinsically non-perturbative, its reliable modeling is essential.
We follow Ref.~\cite{Alameddine:2023wrp} and employ the  parametrization in Ref.~\cite{Abt:2017nkc}, which provides a single parametrization for the entire $Q^2$ range described above. Its lower cutoff is given by the exchange of a pion at rest, $y_{\rm min} \simeq m_\pi/E_\ell$. Similarly to Bremsstrahlung, the energy loss spectrum behaves as $1/y$, possibly jeopardizing the soft expansion in \cref{eq:fokker-planck}. PN is affected by sizeable theoretical uncertainties, especially from the poorly controlled low-$Q^2$ regime at lepton energies above the TeV scale. Different parametrizations lead to different energy-loss rates, and additional uncertainties arise at ultra-high energies from the treatment of atomic shadowing~\cite{KOEHNE20132070}. We therefore use the  parametrization in Ref.~\cite{Abt:2017nkc} as our central value and assign a conservative, energy-independent $30\%$ uncertainty to the PN rate, encompassing both shadowing effects and the spread among parametrizations. This uncertainty affects majorly $d_\mu$, for which PN contributes about $50\%$ of the total for the whole energy range, and is less relevant for $b_\mu$, for which the PN contribution is $25\%$ at $1\,\rm PeV$ and only grows at higher energies, becoming of order $35\%$ at $100\,\rm PeV$.

\begin{table}[t]
    \centering 
    \caption{Relevant quantities describing the propagation of $1~\PeV$ and $100~\PeV$ muons in water. 
    }\label{tab:b-d}
\begin{tabular}{c c c c c}
\toprule
Energy & \multicolumn{2}{c}{$1\,\PeV$} & \multicolumn{2}{c}{$100\,\PeV$} \\
\midrule
Process & $b_{\mu}/{\rm km^{-1}}$ & $d_{\mu}/{\rm km^{-1}}$ & $b_{\mu}/{\rm km^{-1}}$ & $d_{\mu}/{\rm km^{-1}}$ \\
\midrule
$e^{\pm}$ pair production&$0.17$&$4.01 \cdot 10^{-3}$&$0.17$&$4.01 \cdot 10^{-3}$\\
Bremsstrahlung&$9.06 \cdot 10^{-2}$&$4.28 \cdot 10^{-2}$&$9.07 \cdot 10^{-2}$&$4.28 \cdot 10^{-2}$\\
Photonuclear interactions&$8.84 \cdot 10^{-2}$&$2.98 \cdot 10^{-2}$&$0.14$&$5.14 \cdot 10^{-2}$\\
\midrule
Total&$0.35$&$7.66 \cdot 10^{-2}$&$0.40$&$9.82 \cdot 10^{-2}$\\
\bottomrule
\end{tabular}
\label{eq:tabellalosses}
\end{table}
The values in \cref{eq:tabellalosses} clarify the regime of validity of the soft expansion. PP is the softest contribution with $b_\mu/d_\mu\approx2.4\times10^{-2}$. B and PN on the other hand, have $d_\mu/b_\mu\approx0.5$ and $d_\mu/b_\mu\approx0.35$ respectively, on account on their more pronounced hard tails, visible in the left panel of \cref{fig:dsdy}. After summing over all QED processes, we find a global expansion parameter of order $d_\mu/b_\mu\sim 0.2\,$. The soft expansion is therefore not parametrically sharp, but remains marginally convergent. This allows us to keep the expansion in \cref{eq:fokker-planck} as a consistent description of the full QED collisional integral. In practice, the drift is stabilized by the dominance of soft PP, while the diffusion coefficient is more sensitive to the harder B and PN components. We therefore expect corrections beyond the Fokker--Planck approximation at the level of tens of percent.

\subsection{Fitting the transport coefficients}
\label{sec:MuonEnergyLoss:Fits}

In this section we aim to provide realistic prior distributions for the QED parameters, $b_\mu$ and $d_\mu$, that describe the muon propagation in a medium. We quantify these priors by studying propagation in pure water. These values are shown in \cref{eq:tabellalosses}.

To this end, we perform a MC simulation of the muon propagation: at each step $\Delta x_i$, we associate a relative energy $E_i/E_0$, where $E_0$ is the initial muon energy; in the rest of this section, we fix $E_0 = 10$ PeV.
The steps are drawn from an exponential distribution whose expected value is the particle's mean free path (MFP) in the medium. As discussed above, the B cross-section is IR-divergent, hence the notion of MFP is ill-posed without the introduction of an ad-hoc IR cut-off, $y_{\rm min}$. We choose $y_{\rm min} = 10^{-7}$, meaning that we neglect all B processes with $y<y_{\rm min}$, and compute the MFP accordingly.
At each step $\Delta x_i$ the energy is updated by sampling a random value of $y$ according to the QED energy loss spectrum, $\dd\Gamma/\dd y$. The simulation stops when the condition $E_i \leq 0.1\,E_0$ is reached.
To avoid under-sampling of the hard region, we split the PP spectrum in four equally spaced ranges of $y$, from $y = 10^{-4}$ to $y=1$ (we checked that smaller value of $y$ contribute at the sub-percent level to the coefficients $b_\mu$ and $d_\mu$), while we keep the full spectra for B and PN. For each of these, we perform $10^4$ separate MC simulations, for a total of $5\times10^4$ simulations. We use these runs as the data from pseudo-experiments, and fit models of lepton energy loss to them.

Following the discussion in \cref{sec:TransportEquation}, we consider two reference models for the lepton propagation: i) exponential energy loss, corresponding to the drift limit in \cref{eq:Greens:drift}, ii)  drift-diffusion in energy, corresponding to the solution in \cref{eq:Greens}. In both cases, we compute the posterior distributions of the relevant model parameters, $b_\mu$ and $\{b_\mu,d_\mu\}$, with respect to the expected theoretical value reported in \cref{tab:b-d}. The shape of these distributions marks the systematic uncertainty stemming from the propagation model. 
Finally, we include systematics from the uncertainties of the PN low-$Q^2$ region by repeating the fits with the value of the PN rate rescaled by $\pm30\%$, in order to gauge the overall effect on the drift and diffusion coefficients.
The resulting distributions will then be used as informed prior distributions for the parameters when fitting experimental data at neutrino telescopes, see \cref{sec:FitToData}. 

In the following we describe the procedure in the two models.

\begin{figure}[t]
    \centering
    \includegraphics[width=0.45\linewidth]{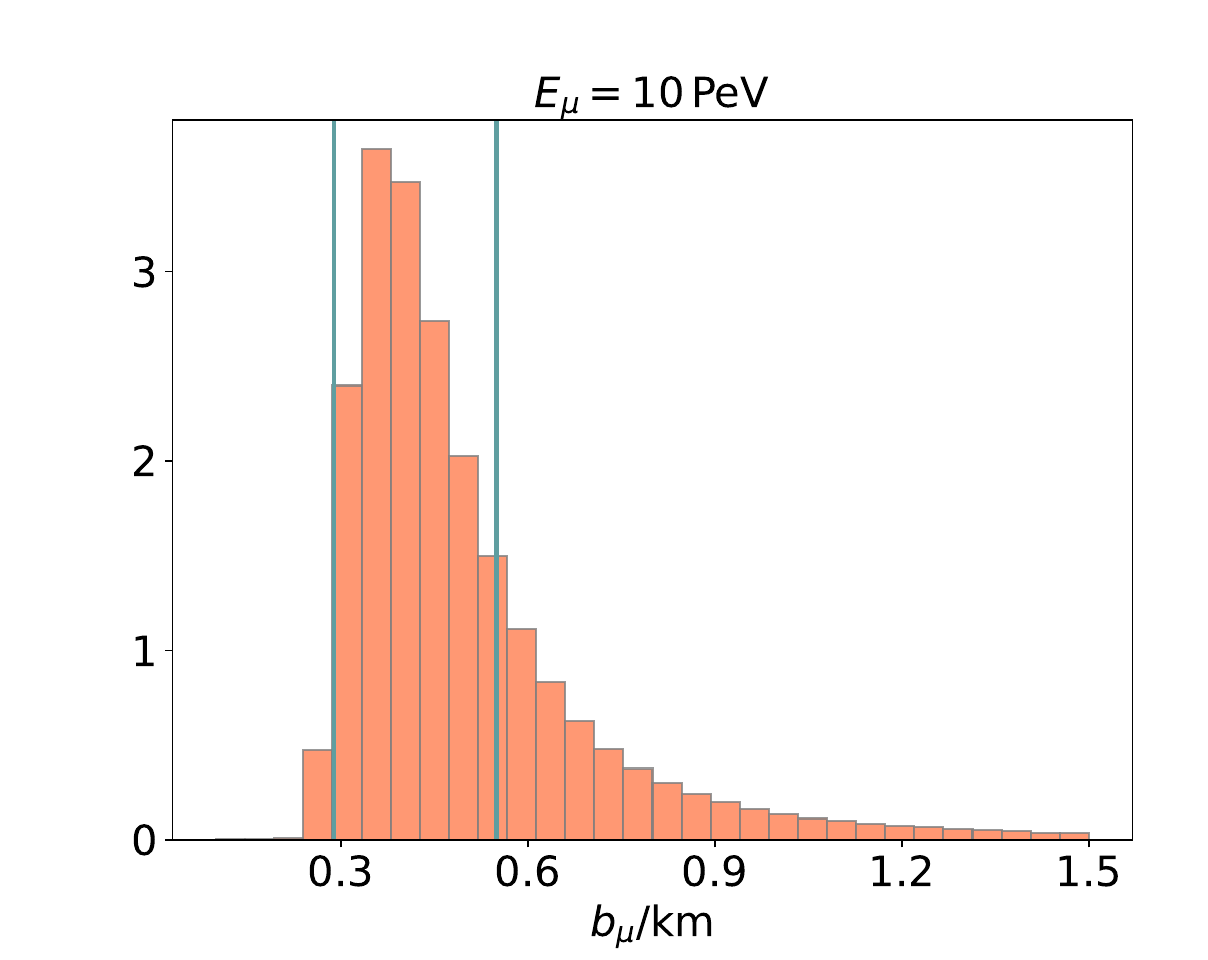}
    \includegraphics[width=0.45\linewidth]{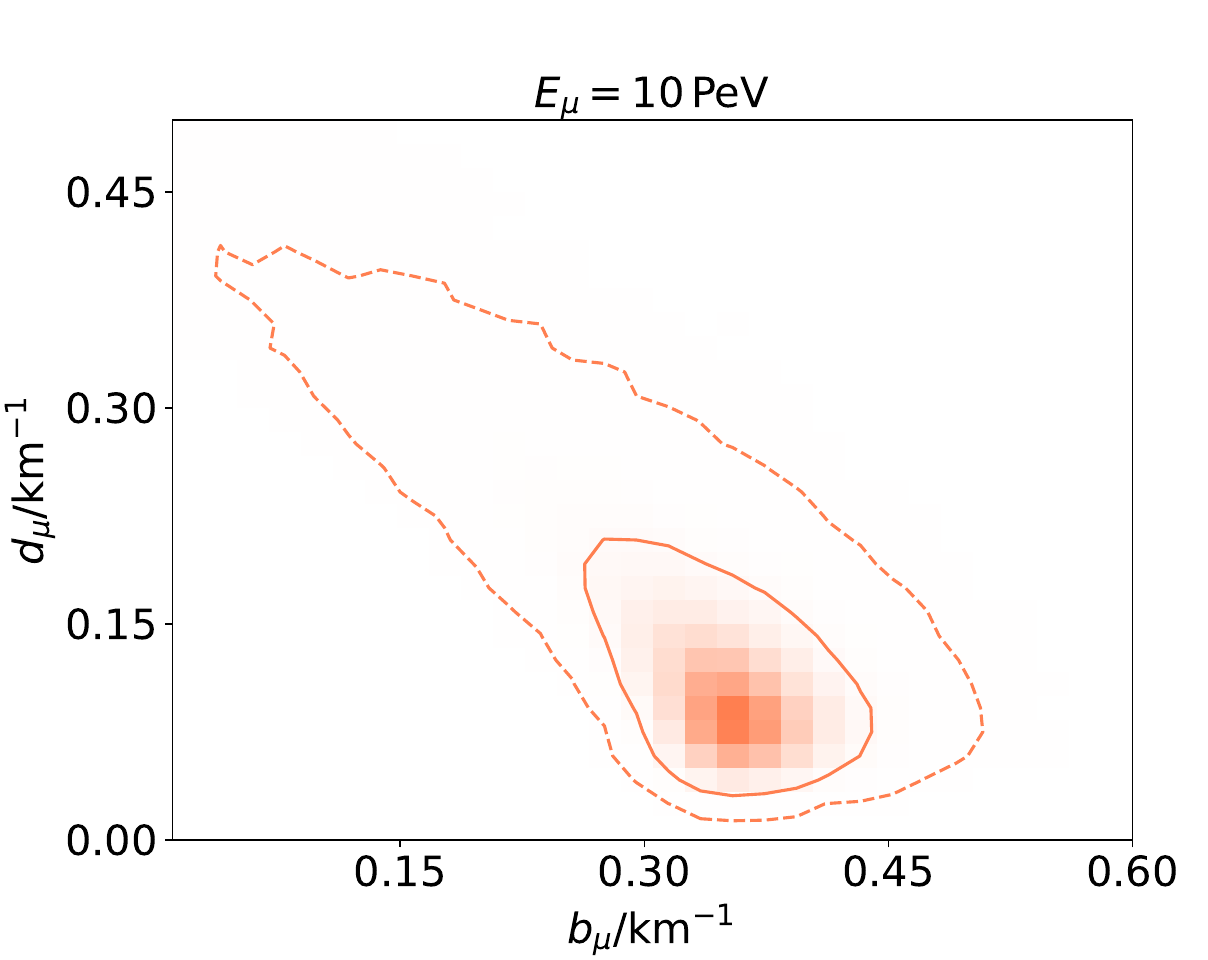}
    \caption{Results of the MC. {\bf Left:} Distribution of the muon drift coefficient $b_{\mu}$. The vertical lines show the highest density 68\% interval. {\bf Right:} 68\% (solid) and 95\% (dashed) highest density regions of the joint distribution of the muon drift and diffusion coefficients, $b_{\mu}$ and $d_{\mu}$.}
    \label{fig:bmuDist_Drift}
\end{figure}

\paragraph{Drift model}
In the drift model, leptons loses energy exponentially due to the QED-induced drift. Hence the expected distance traveled by a muon in a specific medium is determined uniquely by its initial and final energies,  $E_0$ and $E_f$, as
\beq
\langle R \rangle = -\frac1{b_\mu}\log\lp \frac{E_f}{E_0} \rp\,.
\eeq
At fixed $E_0$ there is a unique relation between the expected range at energy $E_f$ and the drift coefficient, which can be inverted to obtain the distribution of $b_\mu^{\rm{MC}}/b_{\mu}$. We find
\beq\label{eq:BestFits:Drift}
b_\mu^{\rm{MC}}/b_{\mu} = 1.03_{-0.22}^{+0.50} ~ \lp\pm~ 0.08\rp\,,
\eeq
where $b_\mu$ at the denominator is the theoretical value, computed for $E_\mu = 10$ PeV. The central value is where the distribution is maximum, and the uncertainty corresponds the 68\% highest-density interval. The error in parenthesis indicates the shift due to the uncertainty on the PN total rate; the total uncertainty on $b_\mu^{\rm{MC}}/b_{\mu}$ can be obtained by summing in quadrature the fit and theoretical errors. In the drift model, the latter is subdominant with respect to the former.

In the left panel of \cref{fig:bmuDist_Drift} we show the resulting distribution, with the central value rescaled to coincide with $b_\mu$. This result indicates that the drift coefficient extracted from the MC simulations of the muon propagation is compatible with the theoretical expectation, although with a sizable uncertainty. The latter will result in a $\mathcal{O}(1)$ uncertainty in the predicted number of events at neutrino experiments.

\paragraph{Drift-diffusion model}

In the drift-diffusion model, the lepton energy distribution at a given traveled distance follows the log-normal distribution in \cref{eq:Greens}. The drift in energy is controlled by the coefficient $b_\mu$, with a small correction from the diffusion coefficient $d_\mu$, which drives the spread of the distribution. 

It is convenient to transform the distribution of $E$, for fixed $\varepsilon$ and $x-\xi$, as a function of $z = \log(E/\epsilon)$, defined by
\beq
\int_0^\epsilon G\lp E, x;\varepsilon, 0 \rp \dd E = \int_{-\infty}^0 {\cal G}\lp z|x \rp \dd z\,.
\eeq
The distribution reads
\beq
{\cal G}\lp z|x \rp = \sqrt{\frac{2}{\pi d_\mu x}}\exp\left[ -\frac{\lp z + M_\mu x \rp^2}{2 d_\mu x}\right]\,,
\eeq
where $M_\mu = b_\mu + d_\mu/2$. With this change of variable, the Green's function resembles a normal distribution with varying width.

For a set of $N$ pairs $\{x_i,z_i\}$, measured from as many MC pseudo-experiments, we consider the following likelihood
\beq
{\cal L} = \prod_i^N {\cal G}\lp z_i|x_i \rp\,.
\eeq
The values $\hat M_\mu$ and $\hat d_\mu$ of the parameters where it reaches its maximum can be derived analytically as
\beq
\hat M_\mu = -\frac{\sum_i z_i}{\sum_i x_i}\,, \qquad \hat d_\mu = \frac{1}{N}\sum_i\frac{\lp z_i + \hat M_\mu x_i \rp^2}{x_i}\,.
\eeq
An anti-correlation between the drift and diffusion coefficients is manifest.
One would expect to observe it in the fit of the MC pseudo-experiments to the full distribution ${\cal G}$. However, this procedure is plagued by a very high computational cost, as the hard part of the QED spectrum is difficult to properly sample, behaving as $1/y$. One way to compensate for under-sampling the hard region is to split the spectrum in smaller intervals of $y$ and perform the fit separately. The two-parameter fit required for the drift-diffusion model would nonetheless be underestimating the values of the coefficients, preferring softer regimes. 

In practice, an easier way to obtain the joint distribution of $b_\mu$ and $d_\mu$ is to solve for the moments, instead of the full distribution, as we already did in the drift case in the previous section. Namely, we can use again the observable $R$ and compute its first two central moments on the distribution ${\cal G}$:
\begin{align}
\langle R \rangle &= \frac{z}{M_\mu} + \frac{d}{M_\mu^2}\,, \\
\langle R^2 \rangle - \langle R \rangle^2 \equiv \sigma_R^2 &= \frac{z~d}{M_\mu^3} + \frac{2d^2}{M_\mu^4}\,.
\end{align}
This system of equations is closed and can be solved to extract $M_\mu = b_\mu + d_\mu/2$ and $d_\mu$ from each measured value of $\langle R \rangle$ and $\sigma_R^2$. To obtain sufficient statistics, we split the full set of MC runs in smaller batches of 10 events each. Following this procedure we find
\beq
b_\mu^{\rm{MC}}/b_\mu = 0.94 \pm 0.15 ~\lp \pm ~ 0.08 \rp\,, \qquad d_\mu^{\rm{MC}}/d_\mu = 1.5_{-0.8}^{+1.6}~\lp \pm ~ 0.15 \rp\,,
\eeq
where $b_\mu$ and $d_\mu$ in the denominator denote the theoretical values, computed for $E_\mu = 10\,\PeV$ (recall that they have a mild dependence on energy). The central values and errors correspond, respectively, to the maxima and the $68\%$ highest-density intervals of the marginal distributions.
The values in parenthesis show again the systematic induced by the uncertainties on the PN rate. The error on $b_\mu$ is the same as for the drift model, albeit it is more relevant: the sum in quadrature brings the total error to 17\%. The effect on $d_\mu$ is larger as expected, although it is negligible with respect to the the fit confidence interval. The rescaled joint distribution is shown in the right panel of \cref{fig:bmuDist_Drift}.

The inferred drift coefficient remains close to the theoretical prediction and is significantly better constrained than in \cref{eq:BestFits:Drift}. The diffusion coefficient is also consistent with the theoretical value, although with a large, asymmetric uncertainty. This behavior is expected: $b_\mu$ is controlled by the first moment of the energy-loss distribution, whereas $d_\mu$ probes the second moment and is therefore more sensitive to rare large-loss events. Equivalently, the statistical uncertainty on the extraction of $d_\mu$ is controlled by the hard part of the collisional integral, making it intrinsically less stable than the drift. The fit therefore supports the Fokker--Planck description at the level of the leading drift term, while showing that the diffusion term is the first place where the hard tail of the QED energy-loss distribution becomes numerically important.

The large relative uncertainty on $d_\mu$, however, does not directly translate into a large uncertainty on the predicted number of events. As shown explicitly in \cref{eq:AnalyticSolution:Diffusion}, diffusion corrects the soft volume only through the combination $d_\mu/2b_\mu\approx 0.1$. Thus even an order-one uncertainty on $d_\mu$ affects only a subleading correction to the predicted soft volume.

In principle this result could be further refined by adding a template in the fit for the energy losses induced by hard scatterings. Even though we do not pursue this road here we provide the relevant formulas for this template in \cref{app:HardScatterings}.

\section{Comparison with neutrino telescope data}
\label{sec:FitToData}

As an application of our formalism, we perform a fit to two datasets: IceCube's 9.5 years muon tracks \cite{Abbasi:2021qfz}, and the KM3NeT ultra-high energy event \cite{KM3NeT:2025npi}. The aim of this section is to test the impact of transport systematics on the measurement of neutrino fluxes and, particularly, on the tension between \kmn{} and \ic{} regarding the event \eventname.

\subsection{Power-law diffuse flux at \ic{}}
\label{sec:ic}

\begin{figure}[t]
    \centering
    \includegraphics[width=0.85\linewidth]{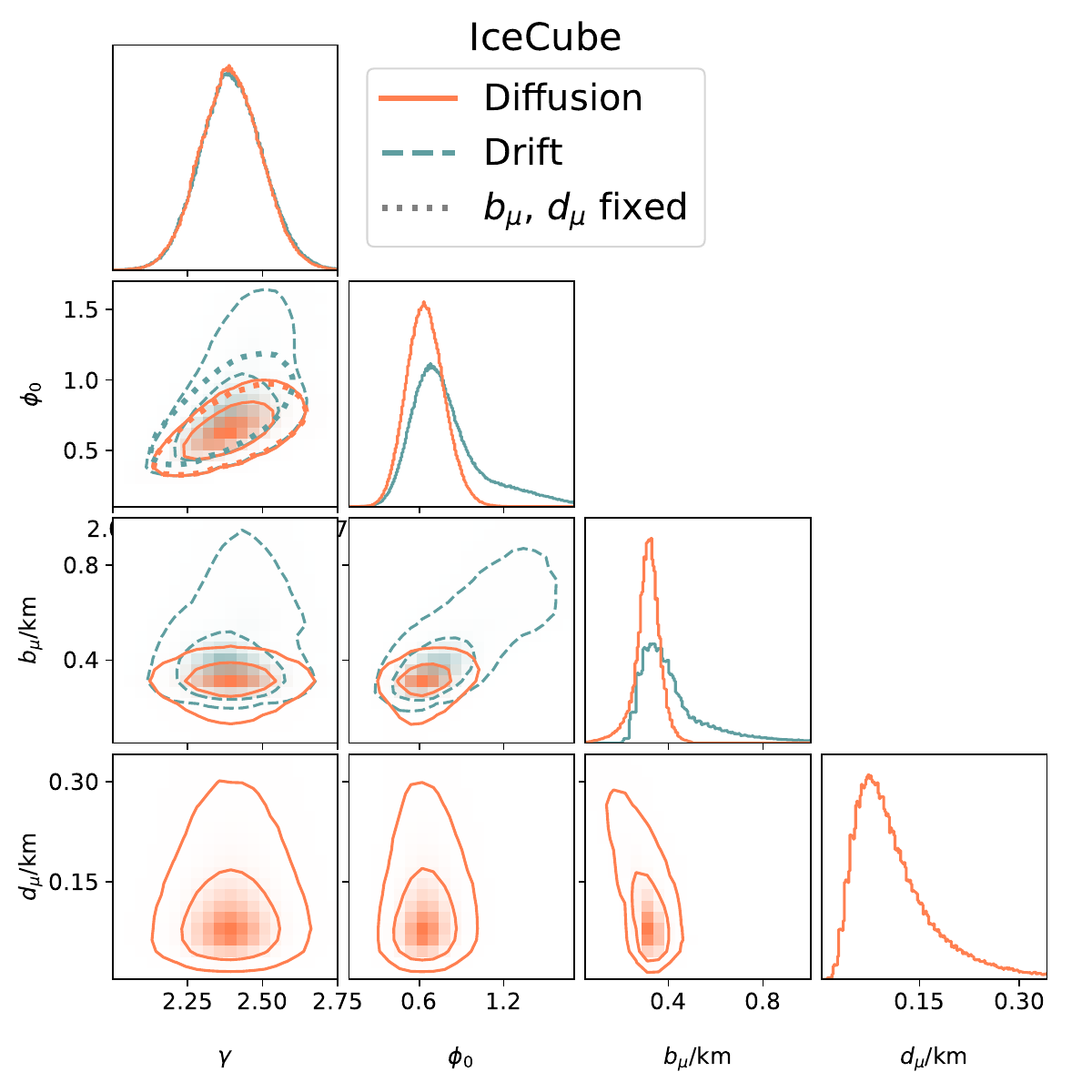}
    \caption{Contours encompassing the 68\% and 95\% of the joint posterior distribution for $\gamma$, $\phi_0$, and $b_\mu$ and $d_\mu$ for IceCube 9.5 years data \cite{Abbasi:2021qfz}, computed in the diffusion model (solid orange) and drift model (dashed blue). Dotted contours are 95\% contours of the posterior assuming nuisance parameters are equal with infinite precision to the values in \cref{tab:b-d}.}
    \label{fig:corner}
\end{figure}

As a first application, we revisit the IceCube measurement of the diffuse astrophysical neutrino flux with through-going muon events~\cite{Abbasi:2021qfz} and quantify how its inferred parameters depend on the theory priors controlling muon propagation, in particular the diffusion parameters entering the soft volume calculation. This exercise provides a direct test of the impact of transport systematics on the flux measurement. At the same time, it illustrates one of the main advantages of our semi-analytical approach, that is, once the propagation kernel is constructed, the theory parameters can be very efficiently varied and marginalized over. We consider a power-law decaying diffuse flux of muon neutrinos,
\beq\label{eq:diffuse-definition}
\phi_\nu^\oplus(E_\nu, \Omega) = \left(\frac{\phi_0}{10^{18}\,\GeV^{-1}\,{\rm cm}^{-2}\,{\rm s}^{-1}\,{\rm sr}^{-1}}\right) \left(\frac{E_\nu}{100\,\TeV}\right)^{-\gamma}\,,
\eeq
which depends explicitly on two parameters, $\phi_0$ and $\gamma$. The overall normalization of $\phi_0$ has been chosen such that, for typical diffuse neutrino fluxes at energies larger than a few tens of $\TeV$, $\phi_0$ is of order unity. Following the prescription in \cite{Abbasi:2021qfz}, we consider all bins with energy above $10\,\TeV$, for a total of 21 bins. For each one, we assume a Poissonian distribution of the events, so that the total likelihood is
\beq\label{eq:likelihood}
\mathcal{L} = \prod_i\frac{\mathcal{N}_i^{k_i} e^{-\mathcal{N}_i}}{k_i!}\,,
\eeq
with $i$ running over the energy bins, $k_i$ being the observed number of muon events, and $\mathcal{N}_i$ being the theoretical expectation. The theoretical expectation includes the expected number of signal events from the astrophysics flux, as well as the expected background noise. The former is a function of the model parameters and is computed as described in the previous sections, while the latter we take in each bin to be equal to the value reported in \cite{Abbasi:2021qfz}.
We consider a uniform prior distribution for the flux parameters $\phi_0$ and $\gamma$. Finally, depending on the model employed for the predictions, we consider two possible nuisance parameters, $b_\mu$ and $d_\mu$, for which we use the posterior distributions found in \cref{sec:TransportEquation} as their prior distributions in the inference procedure.

\begin{table}[t]
    \centering
    \caption{Values of $\phi_0$ and $\gamma$ maximizing the posterior distribution using the various methods described in the previous section, assuming a single power-law diffuse spectrum of neutrinos. The errors correspond to the highest density 68\% intervals. The  approximations refers to those outlined in \cref{sec:soft-volume}.     }\label{tab:fit-results}
    \vspace{5pt}
    \begin{tabular}{c c c}
    \toprule
         Model & $\phi_0/\left(10^{-18} \,\GeV\, {\rm cm}^2\, {\rm sec}\, {\rm sr} \right)$ & $\gamma$  \\
         \midrule
        Diffusion &$0.63^{+0.14}_{-0.13}$&$2.38^{+0.11}_{-0.09}$\\
        Diffusion (approximate) &$0.64^{+0.12}_{-0.13}$&$2.38^{+0.11}_{-0.09}$\\
        \midrule
        Drift &$0.68^{+0.23}_{-0.19}$&$2.39^{+0.11}_{-0.10}$\\
        Drift (approximate) &$0.77^{+0.16}_{-0.15}$&$2.36^{+0.10}_{-0.09}$\\
         \midrule
         \midrule
         IceCube's & $1.44^{+0.25}_{-0.26}$ & $2.36 \pm 0.09 $ \\ 
         \bottomrule
    \end{tabular}
\end{table}

\Cref{fig:corner} shows the 68\% and 95\% credible regions of the joint
posterior distributions for the model parameters obtained from the \ic{}
data of Ref.~\cite{Abbasi:2021qfz}. The corresponding best-fit values are
summarized in \cref{tab:fit-results}. We first observe that the preferred flux normalization, $\phi_0$, is lower
than that reported by the \ic{} Collaboration. As discussed in
\cref{sec:detector}, our ideal-detector approximation overestimates the
number of observed events for a fixed incident flux, thus leading to a smaller inferred normalization. Comparing our result
with that of the \ic{} collaboration, we estimate an effective efficiency
$\epsilon_{\rm IC\text{-}TG}
\equiv\phi_0/\phi_0^{\rm IC}
\simeq 0.45\,,$
as anticipated in \cref{eq:ICeff}.

The magnitude of this suppression can be understood, at least approximately,
by imposing the same cut on the incoming-neutrino zenith angle adopted in
Ref.~\cite{Abbasi:2021qfz}, namely
$\theta_{\rm zenith}>85\degree$. For an isotropic diffuse flux, the fraction
of events surviving this cut is
\begin{equation}
\epsilon_{\rm cut}(E_\nu)
=
\frac{
\displaystyle
\int_{\theta_{\rm zenith}>85\degree}
\dd\Omega\,
D_\nu(E_\nu,\Omega)
}{
\displaystyle
\int_{4\pi}
\dd\Omega\,
D_\nu(E_\nu,\Omega)
}\,,
\end{equation}
where $D_\nu$ is the neutrino attenuation factor defined in
\cref{eq:atten}. We find that $\epsilon_{\rm cut}$ decreases from
approximately $53\%$ at $E_\nu=10\,\TeV$ to approximately $37\%$ at
$E_\nu=1\,\PeV$. The size and energy dependence of this geometrical
acceptance are therefore sufficient to account, at the order-of-magnitude
level, for the difference between our inferred normalization and that
obtained by the collaboration. The actual \ic{} event selection is more involved than the simple
zenith-angle cut considered here and includes additional requirements on
the event topology and reconstruction quality, implemented through
multivariate classifiers trained to reject mis-reconstructed atmospheric
muons~\cite{IceCube:2016umi,Abbasi:2021qfz}.
We therefore do not attempt to model the detector efficiency in greater
detail. Nevertheless, the general event-rate expression in \cref{eq:rate:Complete} remains valid and can be straightforwardly
generalized by introducing an efficiency function, $\epsilon(E_\mu,\Omega)$, parametrizing the probability that a muon of energy $E_\mu$ and
direction $\Omega$ passes the experimental selection. Providing such
efficiency maps would allow the semi-analytic framework developed here to
be applied directly at the detector level.

Secondly, the prior distributions of $b_\mu$ in the drift model and of $d_\mu$ in the diffusion model both exhibit long tails towards large values. In the drift model, the tail of the $b_\mu$ prior induces a corresponding upper tail in the posterior of $\phi_0$, reflecting the strong correlation between the flux normalization and the drift coefficient. This degeneracy persists in the diffusion model, although the narrower prior adopted for $b_\mu$ considerably restricts the allowed range of $\phi_0$. By contrast, we find no significant correlation between $d_\mu$ and either of the flux parameters. Overall, the inferred spectral parameters are remarkably stable: the marginalized posterior distributions of $\gamma$ and $\phi_0$ are only mildly affected by the inclusion of diffusion. The main effect of the additional diffusion parameter is instead confined to the transport sector. In particular, the degeneracy between the drift coefficient and the flux parameters is reduced, while $d_\mu$ itself remains only weakly constrained by the data. This shows that, for the current \ic{} exposure, uncertainties in the soft diffusion coefficient do not constitute the dominant limitation in the reconstruction of the incident neutrino spectrum.

Finally, the dotted lines in \cref{fig:corner} show the $95\%$ credible regions obtained by fixing $b_\mu$ and $d_\mu$ to their theoretical values, thereby treating them as known with infinite precision. The resulting contours for the neutrino-spectrum parameters are only moderately smaller than those obtained after marginalizing over the transport-theory priors in the diffusion model. This indicates that the present uncertainties on $\gamma$ and $\phi_0$ are predominantly statistical and are driven by the finite size of the \ic{} dataset, rather than by uncertainties in the soft muon-transport coefficients.

It is also useful to contrast our ab initio construction with the
conventional parametrization of the through-going event rate, which is
differential in the incoming neutrino energy but inclusive in the muon
energy \cite{Gaisser:2016uoy}. The traditional expression corresponds to an inclusive projection
of the fully differential result, whereas our construction retains the
mapping between the neutrino energy, the muon energy at production, and
the muon energy at the detector. For the diffuse power-law spectrum considered here, the numerical
correction associated with retaining this differential information is
small. At fixed detected muon energy, contributions from parent neutrinos
with $E_\nu\gg E_\mu$ are strongly suppressed by the steeply falling
factor $\phi_\nu(E_\nu)\propto E_\nu^{-\gamma}$, even after accounting for
the growth of the CC cross-section and of the muon range.
Consequently, the convolution is dominated by neutrino energies within a
relatively limited interval above $E_\mu$, and the inclusive
parametrization provides a good approximation. The fully differential
treatment can instead become important for harder or more
ultraviolet-dominated spectra, for which neutrinos at energies far above
the observed muon energy give a parametrically larger contribution.

\begin{figure}
    \centering
    \includegraphics[width=0.7\linewidth]{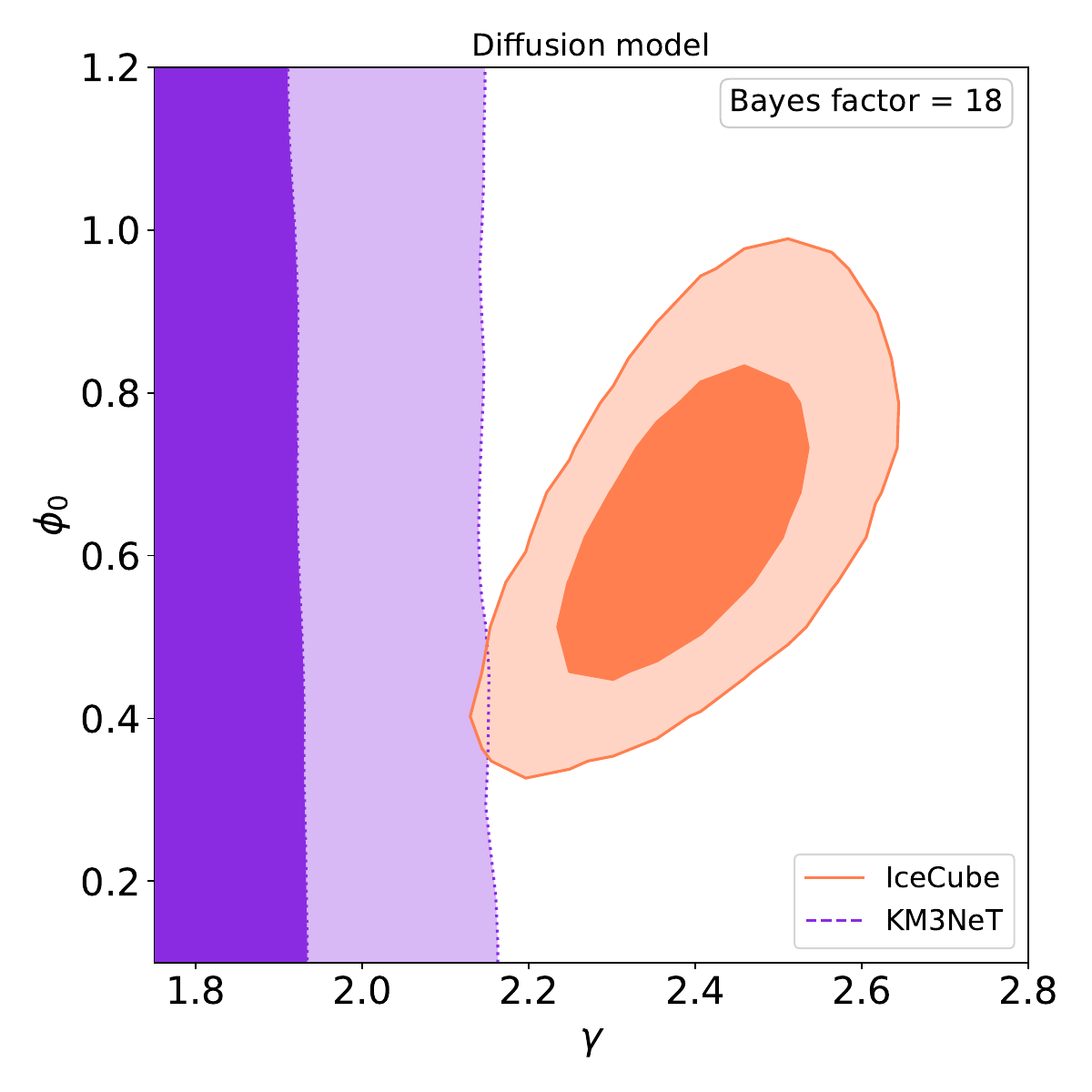}
    \caption{ Best fit results to the diffuse flux parameters $\phi_0$ and $\g$, for the IceCube 9.5 years dataset (orange) and the \eventname~ event (purple). The darker and ligher regions indicate the 68\% and 95\% confidence level regions, respectively. Both fits are obtained considering the diffusion model, resulting in ${\rm BF}=18$, see text for details. }
    \label{fig:ic_v_km3}
\end{figure}

\subsection{The {\eventname} event}
\label{sec:km3event}

The UHE neutrino event \eventname{} was detected by \kmn{} through the
observation of a through-going muon track with reconstructed energy $E_\mu=120^{+110}_{-60}\,\PeV\,,$ where the quoted uncertainties correspond to a 68\% confidence interval;
the corresponding 90\% interval is $E_\mu\in[35,380]\,\PeV$~\cite{KM3NeT:2025npi}.
In the present analysis, we consider a single energy bin containing the
reported muon-energy interval and adopt the likelihood of
\cref{eq:likelihood}, with the product restricted to this bin. We neglect
the atmospheric background at these energies, consistently with the
estimates reported by the \kmn{} collaboration.

A fit of a diffuse power-law flux to a single event inevitably produces a
strong degeneracy between the normalization $\phi_0$ and the spectral
index $\g$, as shown in \cref{fig:ic_v_km3} and previously discussed in
Refs.~\cite{Palmisano:2025abd,titans}. The \kmn{} likelihood favours harder spectra than those inferred by \ic{}, since decreasing $\g$
enhances the UHE tail of the flux. The normalization $\phi_0$ varies
correspondingly along the degenerate direction, such that the expected
number of events in the UHE bin remains of order unity.

We now quantify the compatibility of the \kmn{} event with the diffuse
flux inferred from \ic{}. To this end, we compare two prior distributions
for the same diffuse-flux model. The first, denoted by
$\pi_{\rm flat}$, contains the theory priors on $b_\mu$ and $d_\mu$
described in \cref{sec:QEDlosses}, while adopting flat, broad priors on
$\phi_0$ and $\g$. The second,
\begin{equation}
\pi_{\rm IC}(\hat\theta)
\equiv
\mathcal{P}(\hat\theta\mid\mathcal{D}_{\rm IC})\,,
\end{equation}
is the posterior obtained from the \ic{} analysis and therefore represents
the prior-predictive distribution for the \kmn{} dataset under the
hypothesis that the two experiments observe the same diffuse flux. We
define
\begin{equation}
{\rm BF}_{\rm IC/KM}
=
\frac{
\displaystyle
\int \dd\hat\theta\,
\mathcal{L}_{\rm KM}(\hat\theta)\,
\pi_{\rm flat}(\hat\theta)
}{
\displaystyle
\int \dd\hat\theta\,
\mathcal{L}_{\rm KM}(\hat\theta)\,
\pi_{\rm IC}(\hat\theta)
}\,.
\label{eq:BF-IC-KM}
\end{equation}
A large value of ${\rm BF}_{\rm IC/KM}$ indicates that the \kmn{} event is
substantially better accommodated in the broad parameter space than in
the region selected by \ic{}. In practice, the two evidence integrals can be estimated directly by
Monte Carlo sampling which approaches \cref{eq:BF-IC-KM} in the large-sample limit. Numerically, in the diffusion model we find ${\rm BF}_{\rm IC/KM}\simeq18$
for a prior uniform in $\phi_0$. This result is qualitatively compatible
with Ref.~\cite{Palmisano:2025abd}, where only the drift evolution was
included. According to the conventional Jeffrey's scale, this value constitutes substantial-to-strong evidence that the
\ic{}-informed diffuse-flux model does not provide an adequate description
of the \kmn{} event~\cite{Jeffrey}.\footnote{Although the assumption that the BF follows a chi-square distribution is unjustified given the small amount of data, insisting on it allows one to quantify the tension in the often more familiar terms of $2.8\sigma$.} 

The observation of \eventname{} and its apparent tension with the \ic{} results have motivated a broad range of astrophysical and particle-physics interpretations~\cite{GrimbaumYamamoto:2026kam, KM3NeT:2026flu, Brdar:2025azm, Borah:2025igh,Kohri:2025bsn,Jho:2025gaf, Boccia:2025hpm, Narita:2025udw, Murase:2025uwv, Yuan:2025zwe,SevleMyhr:2026sbk, Alhebsi:2026bdk, Goncalves:2026ofv}. Nevertheless, any explanation involving diffuse or
steadily emitting sources must simultaneously account for the absence of comparable events in the much larger \ic{} exposure. Within the
SM, a transient source remains the most direct way of reducing this tension, because the relevant exposure need not coincide with the
full \ic{} data-taking period~\cite{Palmisano:2025abd,titans}. Letting aside substantial mis-reconstruction of the arrival direction and a possible atmospheric
origin~\cite{Fargion:2025nwl}, another possibility is that \eventname{} represents a rare upward fluctuation. In that case, the rate inferred from \kmn{} will decrease as the experiment
accumulates additional exposure without observing another event of comparable energy, although the evolution of the tension also depends on the concomitant data-taking of \ic{}.

Assuming that is the case, one could quantify what the \kmn{} event implies for the SM parameters, without trying to fix its tension with \ic{}. Ref.~\cite{Bertolez-Martinez:2026bzj} used the nearly--horizontal arrival direction of \eventname{} to constrain the total neutrino--nucleon cross
section. Increasing the total cross section suppresses horizontal and up-going events through neutrino absorption, shifting the normalized
angular distribution towards down-going directions. From the single observed direction, Ref.~\cite{Bertolez-Martinez:2026bzj} obtained
$\sigma_{\nu N}^{\rm tot}
\lesssim
40\,\sigma_{\nu N}^{\rm SM}$
at 95\% confidence level. Because this constraint is extracted from the normalized angular distribution, it is largely insensitive to the
overall normalization of the UHE neutrino flux.

\begin{figure}
    \centering
    \includegraphics[width=0.45\linewidth]{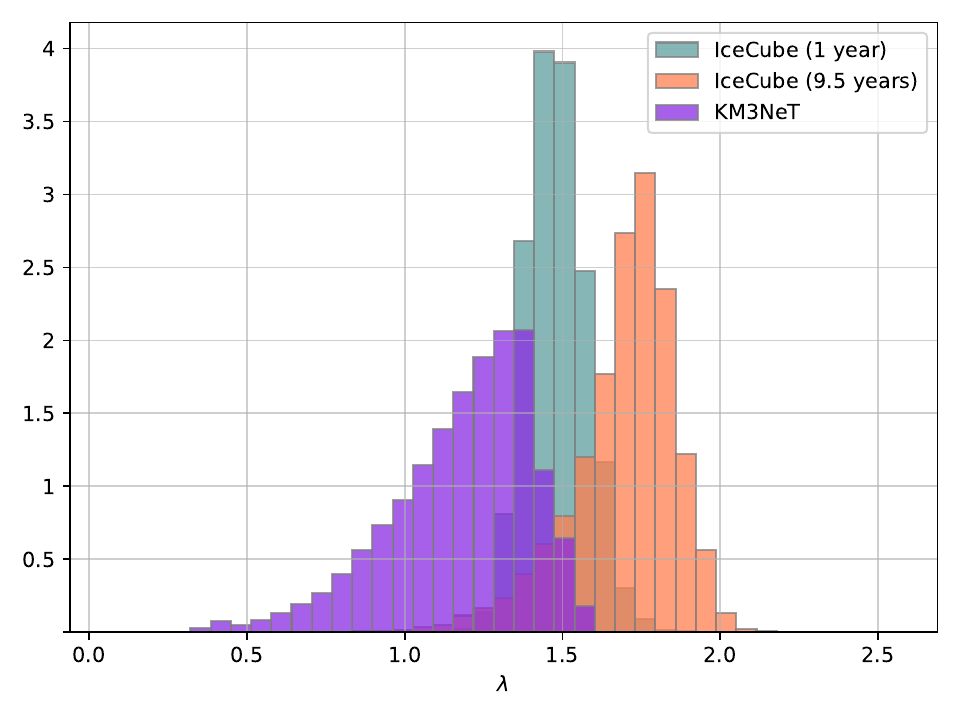}
    \includegraphics[width=0.45\linewidth]{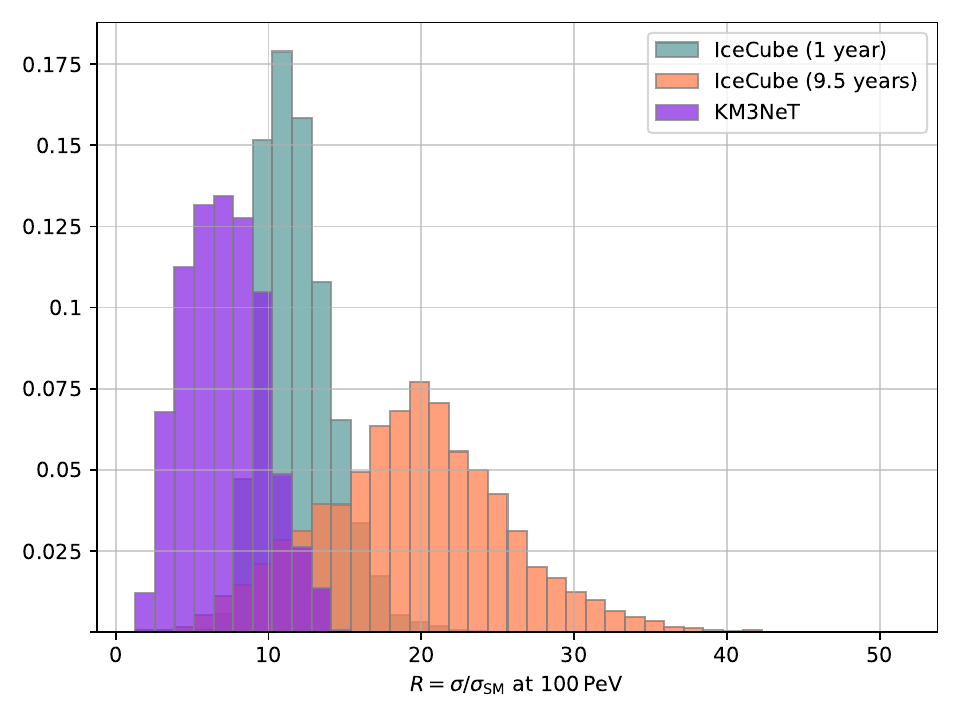}
    \caption{Neutrino cross section required to achieve one muon event at 100 PeV, as computed from \cref{eq:NmuAtEstar}, for KM3NeT full exposure (purple), IceCube with 1 year (green) and 9.5 year (orange) exposure, respectively. {\bf Left:} distribution of the phenomenological parameter $\lambda$. {\bf Right:} required enhancement of the cross section with respect to the SM reference value, computed with $\lambda = 0.4$. }
    \label{fig:Sigma_nuN_Preferred}
\end{figure}

Our formalism provides complementary information through the total event
yield. Within the approximations described in \cref{sec:soft-volume}, and
suppressing factors of angular acceptance, attenuation, and detector efficiency, the number of expected through-going events in a logarithmic
energy interval centred at $E^*$ can be estimated as
\begin{equation}
N_\mu(E^*)
\sim
\frac{\pi R_{\rm det}^2\,n_N}
{b_\mu(\g-\lambda-1)}
\,
\phi_0\,\sigma_0
\left(\frac{E^*}{10\,\PeV}\right)^\lambda
\left(\frac{E^*}{100\,\TeV}\right)^{-\g}
E^*\,T\,,
\label{eq:NmuAtEstar}
\end{equation}
where $T$ is the detector exposure time.
Requiring $N_\mu = 1$ at $E^*=100$ PeV, we can extract the preferred values of $\lambda$, by considering the values of $\phi_0$ and $\g$ preferred by each experiment, as inferred in the previous sections. The preferred values of $\lambda$ for the cases of KM3NeT, IceCube with 1 year and 9.5 years exposure, are presented in the left plot of \cref{fig:Sigma_nuN_Preferred}; all distributions peak around $\lambda\sim1-1.5$, independently of the flux parameters, which would correspond to $R\equiv\sigma/\sigma_{\rm SM}\sim10$ at $100\,\PeV$, where we take for the SM cross section, $\sigma_{\rm SM}$, a reference value $\lambda = 0.4$~\cite{Cooper-Sarkar:2011jtt}. In details, we get 
$\lambda_{\rm KM3Net} = 1.24_{-0.6}^{+0.25}\,,~~  \lambda_{\rm IC,1year} = 1.48_{-0.24}^{+0.12}\,,~~ \lambda_{\rm IC,9.5year} = 1.66_{-0.4}^{+0.26}\,,$ which correspond to $R_{\rm KM} = 6_{-4}^{+5}\,,~~ R_{\rm IC,1year} = 9.8_{-3}^{+6}\,,~~ R_{\rm IC,9.5year} = 10_{-5}^{+20}\,.$

The parameter $\lambda$ should be interpreted only as an effective slope over the finite energy interval probed here. Extrapolating a power-law growth with $\lambda>0$ to arbitrarily high energies would eventually be incompatible with the asymptotic Froissart behaviour of total cross sections~\cite{Froissart:1961ux}. If the required enhancement is attributed to the small-$x$ evolution of the nucleon PDFs within Standard Model DIS, the values $\lambda\sim1$--$1.5$ would be considerably steeper than conventional extrapolations based on HERA data~\cite{Cooper-Sarkar:2011jhk,Cooper-Sarkar:2011jtt}. Conversely, if the enhancement originates from new physics, $\lambda$ can no longer be identified directly with a PDF exponent. Importantly, the very large incident neutrino energy does not imply a comparably large scale for the underlying new dynamics. The relevant partonic center-of-mass energy is
$\sqrt{\hat s}\simeq \sqrt{2x m_N E_\nu}\,,$
while the weak-boson propagator selects momentum transfers $Q^2\simeq2x m_NE_\nu y\sim M_W^2$. Hence,
$\sqrt{\hat s}\sim M_W/\sqrt{y}$ which lies parametrically near the EW scale for the inelasticities relevant to UHE neutrino scattering. New physics capable of modifying the cross section may therefore involve degrees of freedom at scales of order the electroweak scale, rather than at the full incident neutrino energy. Such dynamics would nevertheless have to produce a substantial enhancement of the UHE neutrino cross section while remaining compatible with collider and DIS constraints. Moreover, the growth must eventually soften or become unitarized at higher energies. Satisfying all these requirements simultaneously makes the new-physics interpretation highly constrained.

\section{Conclusions and Outlook}
\label{sec:conclusions}

In this paper, we have developed a streamlined framework for mapping an incident neutrino flux onto the expected muon event rate at high-energy neutrino detectors. For an arbitrary neutrino flux, the formalism yields the number of muon tracks reaching the detector, differential in both the muon energy and the incoming direction. The event rate naturally separates into two contributions. The first is the detector-volume contribution, arising from muons produced within the instrumented volume. The second is the soft-volume contribution, due to muons produced outside the detector that subsequently propagate into it. This contribution depends on muon energy losses in the surrounding medium, which we describe analytically by expanding the QED collision operator in its soft moments. The expectation values of these moments can be computed analytically, while their probability distributions are calibrated against QED simulations of muon propagation. Crucially, this calibration needs to be performed only once, after which the resulting parametrization permits a fast marginalization over the QED energy-loss parameters.

The general expression for the event rate still involves a small number of numerical integrations. However, the simplicity of the resulting kernels makes it possible to introduce a further approximation that reduces the calculation to a fully analytic form. As we have shown, this approximation retains excellent accuracy over the energy range of interest, while making the dependence on the neutrino flux, the detector geometry, and the muon-propagation parameters particularly transparent. The framework therefore provides both a precise semi-analytic prediction and a compact analytic approximation suitable for rapid phenomenological studies and statistical inference.

As a first application of this map, we consider the IceCube through-going muon dataset~\cite{Abbasi:2021qfz}. Assuming a diffuse impinging neutrino flux parametrized by a single power law, we inferred the flux normalization and spectral index while marginalizing over the QED energy-loss parameters. The spectral index we obtain is compatible with the IceCube result, while comparing the respective results for the normalization allows us to estimate the IceCube efficiency and geometric acceptance at about $50\%$; this estimate is consistent with the commonly used $\theta_{\rm zenith}>85\degree$ angular cut on incoming muon tracks, which reject most of the atmospheric background. As a second application, we revisit the tension between the IceCube non-observation and the event observed by KM3NeT~\cite{KM3NeT:2025npi}. We find a BF of $18$, corresponding to a substantial tension. In particular, the present data do not allow one to unambiguously distinguish a genuine inconsistency from a statistical fluctuation of a poorly constrained high-energy flux component. 

Looking ahead, we envision two directions in which the map derived here can be particularly useful. The first is a direct measurement of the soft volume from data. For example, one can compare high-energy through-going events (HETGE)~\cite{IceCube:2016umi,Abbasi:2021aiy}, which are produced outside the detector, with high-energy starting events (HESE)~\cite{IceCube:2020wum}, for which a veto on the outer PMTs selects events whose visible activity starts inside the detector.  Such a measurement would provide an independent test of muon propagation and of the neutrino source flux, giving an additional handle, useful to calibrate the mapping between impinging neutrino fluxes and expected event rates.

The second direction concerns searches for fluxes beyond the diffuse single-power-law component. These may arise from high-energy emission of astrophysical objects, such as blazars~\cite{IceCube:2016qvd,IceCube:2018cha,IceCube:2018dnn,Smith:2020oac}, or from beyond-the-Standard-Model physics, such as dark matter decays~\cite{Chianese:2019kyl,Berghaus:2025jwb}. A common limitation in studying such exotic fluxes is that experimental collaborations usually provide their simulation results in terms of an effective area. In this language, the expected event rate, differential in neutrino energy and incoming direction, is obtained schematically as
\beq
\frac{\dd N}{\dd E_\nu \dd \Omega} \propto A_{\rm eff}(E_\nu,\Omega)\,\phi_\nu(E_\nu,\Omega)\,.
\eeq
The main drawback of this approach is that the effective area is inherently inclusive in the final-state muon energy. Predictions for a given signal therefore rely on neutrino energy reconstructions, which in turn assume a specific spectral shape for the incident flux. If the source spectrum differs significantly from a single power law -- as is the case, for instance, for neutrinos arising from dark matter interactions -- data provided in the form of reconstructed neutrino energies cannot be used directly for comparison with predictions. Furthermore, the effective area implicitly assumes SM physics in both the transport of charged particles and in the interactions between neutrinos and nuclei. If one were to introduce new physics affecting either, the corresponding effective area should in principle be recomputed, which can be computationally expensive, especially when one wishes to scan over large portions of parameter spaces of new-physics models.

Our formalism avoids this limitation by keeping the muon energy and direction explicit throughout the propagation and detector-mapping procedure. It provides a flexible and computationally efficient way to predict the muon event distribution induced by a general impinging neutrino flux. This opens the possibility of extracting less biased information on dark matter decay, transient or localized astrophysical sources, and other non-standard high-energy neutrino fluxes. We plan to return to these applications in future work.

\paragraph{Acknowledgments.}
{\small We thank Shirley Li, Filippo Sala, Matteo Borrello and Ludwig Neste for discussions. The work of AT and SP is
supported in part by the Italian Ministry of University and Research (MUR) through the PRIN
2022 project n. 20228WHTYC~(CUP:I53C24002320006). MT
acknowledges support by Next Generation EU, as part of Piano Nazionale di Ripresa e Resilienza
(PNRR), Missione 4, Componente 2, Investimento 1.2 - CUP I13C25000150006. We are grateful to the Galileo Galilei Institute in Florence, where this work was carried out, for its warm hospitality and stimulating scientific environment. This work was performed in part at the Aspen Center for Physics, which is supported by a grant from the Simons Foundation (1161654, Troyer).}
\newpage

\appendix
\section{Soft expansion of the QED collision operator}
\label{app:qed-collisional}

In this appendix we derive the Fokker--Planck equation used in the main text, starting from the QED collisional term of the full transport equation.  We focus on muon propagation, but the discussion applies more generally to ultra-relativistic charged particles, whose electromagnetic energy loss is dominated by soft momentum transfers.

The starting point is the phase-space transport equation
\begin{equation}
\label{eq:app-full-transport}
\left(
\partial_t+\vec v\cdot\nabla_{\vec x}
\right)
f_\mu(t,\vec x,\vec p)
=
C_{\rm QED}[f_\mu](t,\vec x,\vec p)
+\cdots \, ,
\end{equation}
where the dots denote sources, absorption terms, and weak interactions discussed in the main text, see \cref{sec:setup}. In practice we are interested in the stationary regime, where we can also assume $\partial_t f_\mu(t,\vec x,\vec p)=0$. 

The QED collisional term has the schematic gain--loss form
\beq\label{eq:full-QED-collision}
\begin{split}
C_{\rm QED}[f_\mu]=&-\sum_{T,X}\int \dd\Pi_{T|\bar p,X} \frac{1}{2E_p} f_\mu(p) f_T(p_T)|\mathcal M(p\to \bar p+X)|^2(2\pi)^4\delta^4(P)\nonumber\\&+\sum_{T,X}\int \dd\Pi_{T\bar p|X} \frac{1}{2E_p} f_\mu(\bar p) f_T(p_T)|\mathcal M(\bar p\to p+X)|^2(2\pi)^4\delta^4(P)\,,
\end{split}
\eeq
where we defined the total four momentum of the process $P=\bar{p}+p_X-p-p_T$; the sum over $X$ is inclusive in the final state which induces the muon energy loss, while the sum over $T$ is a sum over the scattering targets. The first line is a loss term, as it removes muons from the part of phase space with momentum $p$, while the second line is the gain term, adding muons to it from higher incoming momenta, $\bar p$. Thus, in general the exact QED collisional integral is non-local in the muon momentum.  

We work in the rest frame of the medium,
\begin{equation}f_T(\vec p_T)=(2\pi)^3 n_T \delta^{(3)}(\vec p_T)\, ,
\end{equation}
and trade matrix elements for differential rates. The collisional operator becomes
\beq\label{eq:QED-collision-rate-3d}
C_{\rm QED}[f_\mu](t,\vec x,\vec p)
=
-\int \dd^3\bar p\,
\frac{\dd\Gamma(\vec p\to \vec{\bar p})}{\dd^3\bar p}\,
f_\mu(t,\vec x,\vec p)+
\int \dd^3\bar p\,
\frac{\dd\Gamma(\vec{\bar p}\to \vec p)}{\dd^3  p}\,
f_\mu(t,\vec x,\vec{\bar p}) \,.
\eeq
At high energies, angular deflections induced by QED scatterings are negligible for the present purposes. We therefore approximate the scatterings as collinear, so that the outgoing momentum direction is the same as the incoming one, and the only parameter accounting for the muon energy propagation is its fractional energy loss 
\begin{equation}
y\equiv\frac{E_{\rm in}-E_{\rm out}}{E_{\rm in}}\ .   
\end{equation}
In this approximation, the fully differential transition rate can be written as
\beq\label{eq:dGammad3pFromdy}
\frac{\dd\Gamma(\vec p_{\rm in}\to \vec p_{\rm out})}{\dd^3p_{\rm out}}=\int_0^1 dy
\frac{\dd\Gamma(E_{\rm in},y)}{\dd y}
\delta^{(3)}
\left[
\vec p_{\rm out}-(1-y)\vec p_{\rm in}
\right]\,.
\eeq
Note that in the gain term of \cref{eq:QED-collision-rate-3d}, the role of the incoming and outgoing momenta changes. 
Substituting \cref{eq:dGammad3pFromdy} into \cref{eq:QED-collision-rate-3d} allows us to reduce the QED collisional term to an energy-loss operator along a fixed trajectory ultimately defined by the momentum direction of the parent neutrino source $\hat{p}$. The three-dimensional kernel in \cref{eq:QED-collision-rate-3d} can be then
projected onto a single energy variable, $E$, becoming
\beq\label{eq:CQED-y-f}
C_{\rm QED}[E, \Omega; f_\mu]
=-
\int_0^1 \dd y\,
\frac{\dd \Gamma(E,y)}{\dd y}\,
f_\mu(E,\Omega)+
\int_0^1 \frac{\dd y}{(1-y)^3}
\frac{\dd \Gamma}{\dd y}
\left(
E_y,y
\right)
f_\mu
\left(
E_y,\Omega
\right)\,, 
\eeq
where we used the ultra-relativistic limit, $d^3p = E^2dEd\Omega$, and we defined $E_y=E/(1-y)$; $\Omega\equiv\hat p$ indicates the direction on the parent neutrino. Finally, the factor \((1-y)^{-3}\) is the Jacobian of the three-dimensional map from momenta, $\dd^3\vec p$, to energy loss, $\dd y$. 

To connect with the main text, it is useful to move from the muon distribution to the flux as
\beq\label{eq:FluxVariable}
(2\pi)^3\phi_\mu(E,\Omega)
\equiv
E^2 v f_\mu(E,\Omega)\,.
\eeq
Therefore the projected collision operator, $C_{\rm QED}[E;\phi_\mu]=(2\pi)^3E^2 C_{\rm QED}[E;f_\mu]$, can be written as 
\beq\label{eq:CQEDProjected}
C_{\rm QED}[E;\phi_\mu]=\left[
-\int_0^1 \dd y\frac{\dd\Gamma_i(E,y)}{\dd y}\phi_\mu(E)
+\int_0^1\frac{\dd y}{1-y}\frac{\dd\Gamma\left(E_y,y
\right)}{\dd y}
\phi_\mu
\left(E_y
\right)
\right]\, ,
\eeq
where the factor $(1/(1-y))$ is the remnant of the original three-dimensional Jacobian.

\begin{figure}[t]
    \centering
    \includegraphics[width=0.9\linewidth]{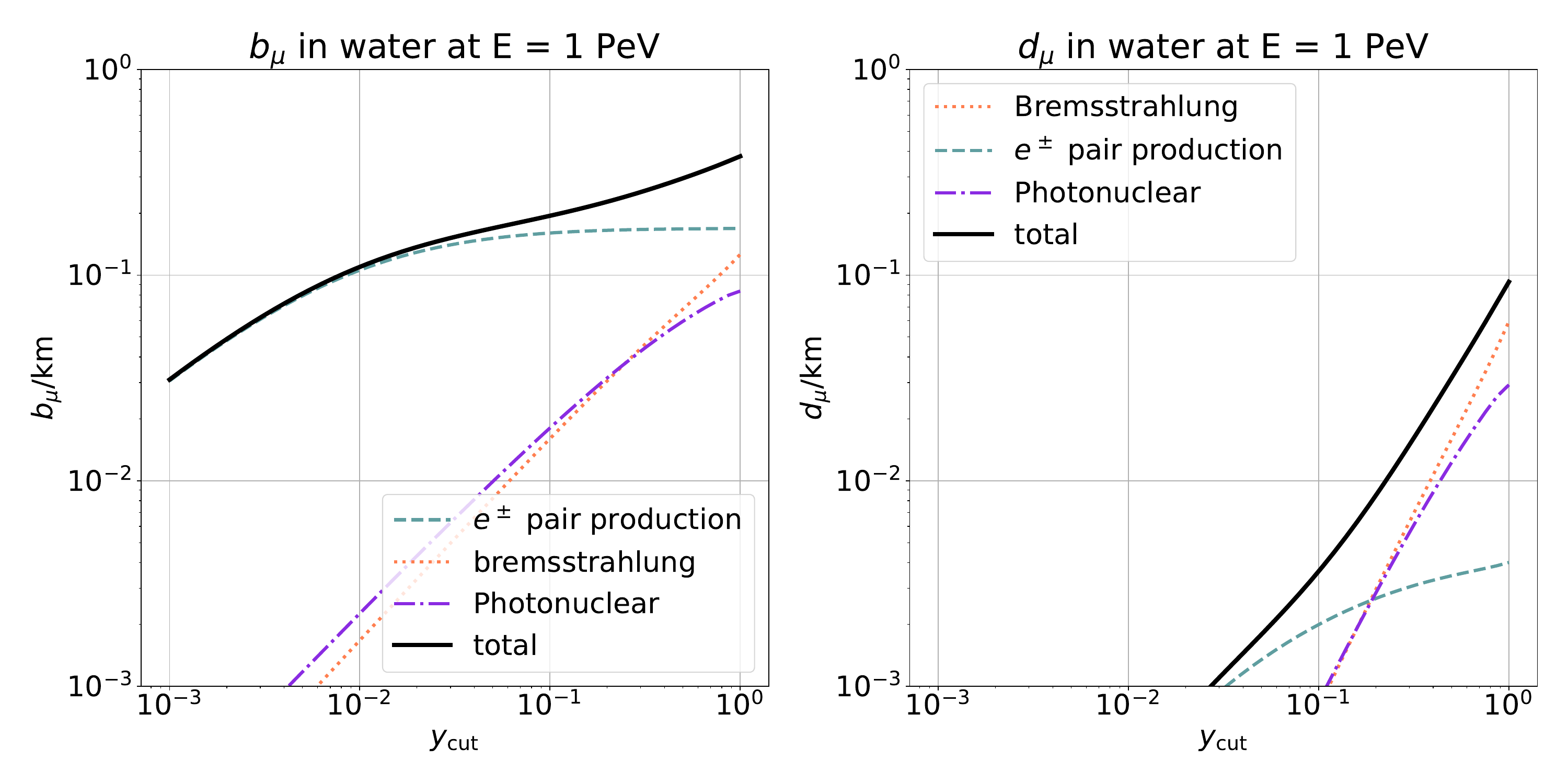}
    \caption{Dependence of $b_\mu$ and $d_\mu$, as defined in \cref{eq:y_moments_cut}, on $y_{\rm cut}$, for muons of $1\,\PeV$ traversing water. Color and style scheme is the same as \cref{fig:dsdy}. }
    \label{fig:b_and_d}
\end{figure}

As discussed in \cref{sec:QEDlosses}, different processes contribute to the energy losses of a high energy muons, which we need to sum over to compute the full form of \cref{eq:CQEDProjected}. Crucially, the different processes have different behaviors as functions of $y$, and different dependences on energy, as discussed in \cref{sec:QEDlosses}. 
Thus, we introduce a cutoff in fractional energy loss, $y_{\rm cut}\lesssim 1$, and split the collision operator into a soft and a hard part,
\begin{equation}\label{eq:softplushard}
C_{\rm QED}[E,\hat p;f_\mu]
=
C_{\rm soft}[E,\hat p;f_\mu]
+
C_{\rm hard}[E,\hat p;f_\mu] \, .
\end{equation}
The soft part is defined by $0<y<y_{\rm cut}$, while the hard part is defined by $y_{\rm cut}<y<1$. For a generic choice of $y_{\rm cut}<1$, the hard contribution remains non-local in energy, and cannot be part of the local Fokker--Planck approximation. Conversely, the soft part of the collisional integral in \cref{eq:softplushard} can be expanded locally, because it samples the distribution only in a small neighborhood of $E$. Expanding for small $y$ gives
\be\label{eq:FP-avatar_app}
C_{\rm QED}[E;\phi_\mu] \simeq
\partial_E
\Big( 
b_\mu(E;y_{\rm cut}) E \phi_\mu(E)
\Big) +
\frac12 \partial_E^2
\Big(
d_\mu(E;y_{\rm cut}) E^2 \phi_\mu(E)
\Big)
\,,
\ee
where we are generalizing the definitions of drift and diffusion coefficients in 
\cref{eq:b-d-def}, to include the dependence on $y_{\rm cut}$ as
\begin{equation} \label{eq:y_moments_cut} 
b_\mu(E;y_{\rm cut}) \equiv 
\int_{y_{\rm min}}^{y_{\rm cut}} \dd y\, y\, \frac{\dd\Gamma(E,y)}{\dd y}\qquad d_\mu(E;y_{\rm cut}) \equiv 
\int_{y_{\rm min}}^{y_{\rm cut}} \dd y\, y\, \frac{\dd\Gamma(E,y)}{\dd y}\,. 
\end{equation} 
In \cref{fig:b_and_d} we show the behaviour of $b_\mu$ and $d_\mu$ as function of $y_{\rm cut}$, for the benchmark case of muon with energy $E_\mu = 1$ PeV propagating in water.
The full process is recovered in the limit $y_{\rm cut}\to1$, as well as the values of the coefficients shown in \cref{eq:tabellalosses}. 

In \cref{sec:setup,sec:FitToData} we kept $y_{\rm cut}\to1$, and assumed that the soft expansion is a good approximation of the full QED collisional integral. As we can see from \cref{fig:b_and_d} we expect this  approximation to be good up to effects of order $\mathcal{O}(d_\mu/b_\mu)\simeq\mathcal{O}(30\%)$. We also assumed $b_\mu$ and $d_\mu$ to be independent on the incoming energy $E$; the latter is a good assumption for PP and B, while it fails for PN at high energies, as shown in \cref{fig:dsdy}. Within these approximations the full QED collisional integral reduces to a Fokker-Planck operator with constant coefficients which we write as
\be
C_{\rm QED}[E;\phi_\mu] \simeq 
\partial_E
\left(
b_\mu E \phi_\mu(E)
\right) +
\frac12 \partial_E^2
\left(
d_\mu E^2 \phi_\mu(E)
\right)
\,.
\ee
At this order, the stationary Boltzmann equation for the flux is
\be
\deriv{\phi_\mu}{x}
-
\deriv{}{E}
\Big(
b_\mu E\phi_\mu(E,x)
\Big)
-
\frac12
\frac{\partial^2}{\partial E^2}
\Big(
d_\mu E^2\phi_\mu(E,x)
\Big)
=
\frac{E^2}{(2\pi)^3}C_{\rm weak}\,,
\eeq
as derived in \cref{eq:fokker-planck}.

The solution of this equation can be easily found with the method of Green's functions, and it is reported in \cref{eq:Greens}. Here we review the crucial steps of this derivation for completeness. The derivation simplifies after changing variable to $u\equiv \log E\,.$ 
Since the Fokker--Planck equation is written for the flux per unit energy, it is convenient to introduce the flux per unit logarithmic energy, $\psi_\mu(u,x)\equiv E\,\phi_\mu(E,x)\,,$ such that  \(\phi_\mu(E,x)dE=\psi_\mu(u,x)du\). 
The homogeneous equation then becomes 
\begin{equation} \label{eq:driftdiff}
\deriv{\psi_\mu}{x} = M_\mu \deriv{\psi_\mu}{u} + \frac{d_\mu}{2} \frac{\partial^2\psi_\mu}{\partial u^2}\,. 
\end{equation} 
Thus the QED energy-loss problem becomes an ordinary drift-diffusion equation in logarithmic energy. The drift velocity in \(u\)-space is $-M_\mu = -\left(b_\mu+\frac{d_\mu}{2}\right)\,$ while the diffusion coefficient is $d_\mu$. 
The Green's function $G_{\rm{soft}}$ for the logarithmic flux is defined by 
\begin{equation} 
\left[ \partial_x - M_\mu\partial_u - \frac{d_\mu}{2}\partial_u^2 \right] G_{\rm{soft}}(u_E,x;u_\varepsilon,x_0) = \delta(x-\xi)\delta(u_E-u_\varepsilon)\,. 
\end{equation}
For propagation length $\ell=x-\xi>0$, the solution is the Gaussian kernel 
\begin{equation} 
G_{\rm{soft}}(u_E,x;u_\varepsilon, \xi) = \Theta(\ell)\, \frac{1}{\sqrt{2\pi d_\mu \ell}} \exp\left[ -\frac{ \left( u_E-u_\varepsilon+M_\mu\ell \right)^2 }{ 2d_\mu \ell } \right]\,,
\end{equation} 
which gives the solution in \cref{eq:Greens}, after going back to the flux per unit energy, and imposing the physical condition that \(\theta(\varepsilon-E)\), so that energy can only be lost in the propagation.  

\subsection{Including the energy dependence of the transport coefficients}
\label{sec:energydep}
In this appendix we discuss how the solution is modified
when the energy dependence of the transport coefficients is retained. We work again in the logarithmic energy variable, \(u\), as defined to derive \cref{eq:driftdiff}.

We assume that $b_\mu(E)$ and $d_\mu(E)$ are slowly varying functions of logarithmic energy,
\beq\label{eq:slowly}
\left|\frac{\partial\log b_\mu}{\partial u}\right|\ll 1,
\qquad
\left|\frac{\partial\log d_\mu}{\partial u}\right|\ll 1.
\eeq
Thus the coefficients may change appreciably over many decades in energy, while still being approximately constant over the width of the Green's function associated with a single propagation interval. The homogeneous part of \cref{eq:fokker-planck} can be written as
\beq
\label{eq:FP-u-variable}
\deriv{\psi_\mu}{x}
=
\deriv{}{u}
\left[
M_\mu(u)\psi_\mu(u,x)
\right]
+
\frac12
\frac{\partial^2}{\partial u^2}
\left[
d_\mu(u)\psi_\mu(u,x)
\right]\,.
\eeq
Let \(\bar u(x)\) be the deterministic trajectory starting from \(u_\varepsilon\) at \(x=\xi\); namely
\beq
\label{eq:ubar-equation}
\frac{\dd \bar u}{\dd x}
=
-M_\mu(\bar u),
\qquad
\bar u(\xi)=u_\varepsilon.
\eeq
Equivalently,
\beq
\label{eq:range-beta-u}
x-\xi
=
\int_{\bar u(x)}^{u_\varepsilon}
\frac{\dd v}{M_\mu(v)}.
\eeq
Since \(b_\mu\) and \(d_\mu\) vary slowly over the logarithmic width explored by the propagator, the Green's function is still approximately Gaussian in \(u\). It follows that its mean is the trajectory \(\bar u(x)\), and
its variance \(\Sigma\) satisfies
\beq
\label{eq:Sigma-equation}
\frac{\dd \Sigma}{\dd x}
=
d_\mu(\bar u)
-
2M_\mu'(\bar u)\Sigma,
\qquad
\Sigma(\xi)=0.
\eeq
The first term is the local diffusion generated along the trajectory. The second term accounts for the compression or stretching of nearby trajectories due to the energy dependence of the drift. The solution for the variance is
\beq
\label{eq:Sigma-solution-x}
\Sigma(x;\xi,u_\varepsilon)
=
\int_\xi^x \dd s\,
d_\mu(\bar u(s))
\exp\left[
-2\int_s^x \dd t\,M_\mu'(\bar u(t))
\right].
\eeq
Equivalently, using \(\bar u\) itself as integration variable,
\beq
\label{eq:Sigma-solution-u}
\Sigma(x;\xi,u_\varepsilon)
=
M_\mu(\bar u(x))^2
\int_{\bar u(x)}^{u_\varepsilon}
\dd v\,
\frac{d_\mu(v)}{M_\mu(v)^3}.
\eeq
Returning to the energy variable, the slowly varying Green's function gives
\beq
\label{eq:Greens-slow-varying}
G_{\rm{soft}}(E,x;\varepsilon,\xi)
=
\frac{\theta(x-\xi)}
{E\sqrt{2\pi\Sigma(x;\xi,u_\varepsilon)}}
\exp\left[
-\frac{
\left(u_E-\bar u(x)\right)^2
}
{2\Sigma(x;\xi,u_\varepsilon)}
\right].
\eeq
As in \cref{eq:Greens}, one may multiply this expression by
\(\theta(\varepsilon-E)\) to enforce the physical support of the underlying energy-loss process. The small Gaussian tail at \(\varepsilon<E\) is an
artifact of the second-order Fokker--Planck approximation.

The solution of the full transport equation is obtained again by convoluting the Green's function with the weak source. The definition of the soft volume that follows remains unchanged: starting from \cref{eq:generalsoft}, one simply replaces the constant-coefficient propagator by \cref{eq:Greens-slow-varying},
\beq
\label{eq:generalsoft-slow-varying}
V_{\rm soft}(E)
=
A_{\rm proj}
\int_0^\infty \dd \xi
\int_E^\infty \dd\varepsilon\,
G_{\rm{soft}}(E,L;\varepsilon,L-\xi)
\left(\frac{\varepsilon}{E}\right)^{-A}\, ,
\eeq
with $A\equiv\gamma-\lambda-1$. Thus the soft volume is affected only through the modified propagation kernel.

In the drift limit we find
\beq
V_{\rm soft}^{\rm drift}(E)
=
A_{\rm proj}
\int_0^\infty \dd\xi\,
\left(
\frac{\varepsilon_\xi(E)}{E}
\right)^{-A}.
\eeq
where \(\varepsilon_\xi(E)\) is the production energy
needed for a muon to arrive at the detector with energy \(E\) after traveling a
distance \(\xi\). Changing variable from \(\xi\) to \(u\), with
\(\dd \xi=\dd u/b_\mu(u)\), gives
\beq
\label{eq:Vsoft-slow-drift}
V_{\rm soft}^{\rm drift}(E)
=
A_{\rm proj}
\int_{u_E}^{\infty}
\frac{\dd u}{b_\mu(u)}
\exp\left[-A(u-u_E)\right].
\eeq
For constant \(b_\mu\), this reduces immediately to
\cref{eq:AnalyticSolution:Drift}. Therefore the slow energy dependence does not change the interpretation of the
soft volume. It remains the projected detector area times the effective muon QED range. The main difference shows up in the latter, which is now a spectrum-weighted average of \(1/b_\mu\) over the energies from which the observed muons can originate.

\section{Including hard collisions}
\label{app:HardScatterings}

In this Appendix we build upon the definition of soft and hard QED collisional terms, \cref{eq:softplushard}, and attempt to improve the description of the QED energy loss  by including the hard piece as a separate perturbative correction to the soft evolution.

The need for an improvement in the computation of the hard collisional operator stems from the discussion in \cref{sec:QEDlosses}, which showed that the description of the full QED energy loss with a template that only contains the drift and diffusion terms leads unavoidably to large theoretical errors on the diffusion. The latter are the manifestation of the fact that the kinematics of B and PN interactions are largely dominated by hard scatterings, $y\sim1$, and by themselves are not correctly described by the soft expansion, unlike the PP case.  
In other words, the large systematics in the soft expansions originate from neglecting the non-local hard operator in \cref{eq:softplushard} for B and PN scatterings, thus an improved description of these terms can reduce the uncertainties.

A systematic way to include the effects of the hard collisional operator is to treat $C_{\rm QED}^{\rm hard}$, defined in \cref{eq:softplushard}, as an insertion on top of the soft evolution. That is, we treat $\phi_0$ perturbatively, such that we can write
$\phi_\mu=\phi_\mu^{(0)}+\phi_\mu^{(1)}+\cdots$, where
$\phi_\mu^{(0)}$ is given in \cref{eq:solution-from-greens}. The leading correction is then obtained through the standard perturbation theory procedure, that is, by assuming that $C_{\rm hard}$ is formally of order $\phi^{(1)}_\mu$. At first order in the perturbative expansion, we obtain the differential equation
\beq
\frac{\dd \phi_\mu^{(1)}}{\dd t} - \frac{E^2}{\lp2\pi\rp^3}C_{\rm QED}^{\rm soft}[\phi_\mu^{(1)}] =  \frac{E^2}{\lp2\pi\rp^3} C_{\rm QED}^{\rm hard}[\phi_\mu^{(0)}]\,.
\eeq
This equation can be solved again by means of the method of Green's functions, by treating $C_{\rm hard}$ as a source. Namely
\beq
\phi_\mu^{(1)}(x,E) = \int\dd \xi\int\dd \varepsilon~ G_{\rm soft}(x,E|\xi,\varepsilon)\frac{\varepsilon^2}{(2\pi)^3}C_{\rm hard}\left[\phi_\mu^{(0)},\xi,\varepsilon,y_{\rm cut}\right]\,.
\eeq
Thus hard scatterings appear as non-local insertions along the soft propagation. This induces a correction to the expected number of events which can be written, in the language of \cref{sec:detector}, as 
\begin{equation}
\left\langle
\frac{\dd N}{\dd t\,\dd E\,\dd\Omega}
\right\rangle\Bigg\vert_{\rm{hard}}=  A_{\rm proj}\phi_\mu^{(1)}(x,E) \,.
\end{equation}
The hard part of the QED collisional integral computed on the zeroth order solution can be explicitly written as
\beq\label{eq:CQED:hard:explicit}
\frac{\varepsilon^2}{\lp2\pi\rp^3}C_{\rm hard}\left[\phi_\mu^{(0)},\xi,\varepsilon,y_{\rm cut}\right] = -\int_{y_{\rm cut}}^1 \!\!\!\!\dd y \frac{\dd\Gamma(\varepsilon)}{\dd y}\phi_\mu^{(0)}(\xi,\varepsilon)+ \int_{y_{\rm cut}}^1 \!\!\frac{\dd y}{1-y} \frac{\dd\Gamma(\varepsilon_y)}{\dd y}\phi_\mu^{(0)}\lp \xi,\varepsilon_y\rp \,,
\eeq
where we defined $\varepsilon_y=\varepsilon/(1-y)$. This is the contribution from $y_{\rm cut}$ to 1 of 
\cref{eq:C_QED_y_general}. The zeroth order term, $\phi_\mu^{(0)}$, and the weak collisional integral, $C_{\rm weak}$, are defined in \cref{eq:solution-from-greens,eq:C_weak}, respectively. Note that this expression for $\phi_\mu^{(1)}$ is general, as no assumption on the form of $G_{\rm soft}$ has been made. Furthermore, the corrections are computed recursively as a perturbation expansion as function of the same $G_{\rm soft}$, thus introducing higher order corrections does not introduce new parameters in the theory, beyond the ones that define $G_{\rm soft}$ itself. In this sense, once the Kramers-Moyal expansion order is fixed, the theory is \emph{renormalizable}. 

\begin{figure}[t]
    \centering
    \includegraphics[width=0.9\linewidth]{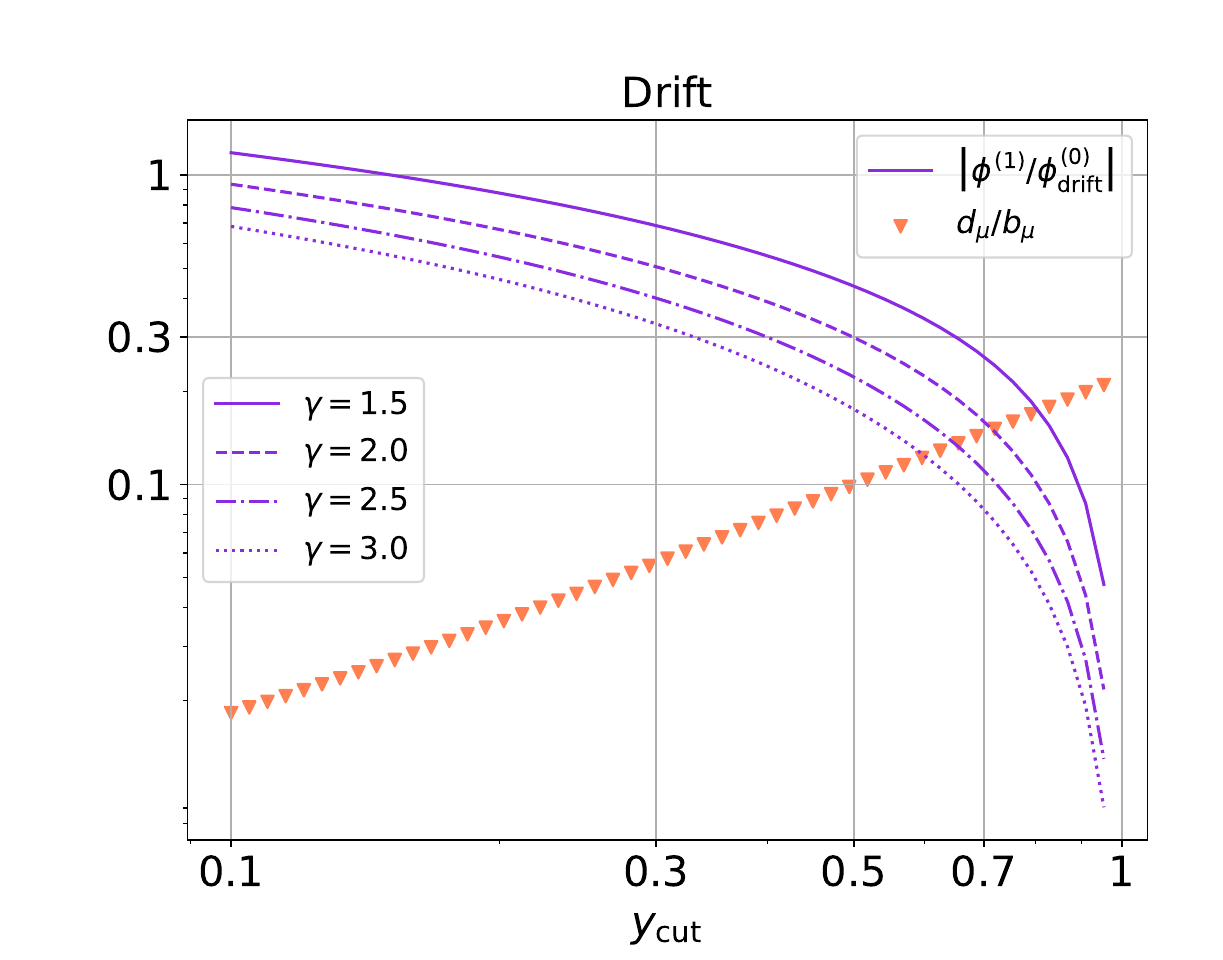}
    \caption{Size of the systematic uncertainties induced by the different treatment of the hard scatterings, as function of $y_{\rm cut}$. The error from the expansion of the full collisional integral, computed as the ratio $d_\mu/b_\mu$, is indicated by the orange triangles, while the estimate of the contribution of hard scatterings with respect to the leading-order soft drift solution, \cref{eq:phi1:driftApprox}, is shown with purple lines, for different values of $\g$.}
    \label{fig:hard-vs-soft}
\end{figure}

As for the leading order result, the integral must in general be computed numerically, but a semi-analytic solution can be found with the same assumptions outlined in \cref{sec:soft-volume}. Both in the drift and diffusion case, the approximate leading-order solution $\phi^{(0)}_\mu$, which can be read off from \cref{eq:AnalyticSolution:Drift,eq:AnalyticSolution:Diffusion}, has only a power-law dependence on energy, inherited from the power-law decaying neutrino flux and from the behaviour of the proton PDF at small Bj\"orken $x$, viz. 
\beq
\label{eq:phi0:Simplified}
\phi_\mu^{
(0)}(E_\mu) = \phi_\nu^\oplus \lp1-\left\langle y_w \right\rangle\rp^{1+A} \lp E_0\rp \lp\frac{ \mathcal{L}}{\lambda_{\nu N}\lp E_0\rp}\rp\lp\frac{E_\mu}{E_0}\rp^{- 1 - A}\,,
\eeq
where $A = \gamma - \lambda - 1$, $E_0$ is a reference energy, $\lambda_{\nu N} =  n_N \sigma_{\nu N}\lp E_0\rp $ is the neutrino mean free path, and $\langle y_w \rangle$ is the average energy loss from neutrino CC interactionst, which we write explicitly with the subscript $w$ to distinguish it from the QED energy loss $y$. The quantity $\mathcal{L}$ is a scale of length which depends on the transport model. We have 
\beq
\begin{aligned}
\mathcal{L} =& \frac{1}{A' b_\mu}\lp1+\sqrt\frac{b_\mu}{B + A' b_\mu}\rp &\text{diffusion}\\
\mathcal{L} =& \frac{1}{A b_\mu} &\text{drift limit}
\end{aligned}
\eeq
where $A'$ and $B^2$ are defined below \cref{eq:AnalyticSolution:Diffusion}. With \cref{eq:phi0:Simplified}, we can now compute $C_{\rm hard}$ as
\beq\label{eq:Chard:Simplified}
\begin{split}
C_{\rm hard}\left[ \phi_\mu^{(0)},\xi,\varepsilon,y_{\rm cut}\right] \approx &\phi_\nu^\oplus\lp E_0\rp\lp\frac{ \mathcal{L}}{\lambda_{\nu N}\lp E_0\rp}\rp\lp\frac{1}{E_0}\rp^{-1-A} \lp1-\left\langle y_w \right\rangle\rp^{1+A} \\
& \int_{y_{\rm cut}}^1\dd y\Bigg[ - \,\varepsilon ^{\lambda-\g} \frac{\dd\Gamma(\varepsilon)}{\dd y}
+  \lp  \frac{\varepsilon}{1-y} \rp^{\lambda-\g} \frac{1}{1-y}\frac{\dd\Gamma(\varepsilon_y)}{\dd y}\Bigg]\,.
\end{split}
\eeq
 These integrals can be computed numerically, but, again, one can estimate them by neglecting the dependency of QED rates on the energy of the processes. As shown in the bottom panels of \cref{fig:dsdy}, this is a reasonable assumption for a large range of energies. Of course, the $1/(1-y)$ enhancement makes the energy blow up as $y$ approaches $1$, but the power-law decaying flux effectively cuts off these contributions (recall that $\lambda - \gamma < 0$). The integral over $\varepsilon$ is then the same that one evaluates to compute $\phi^{(0)}_\mu$, so we can write
\beq\label{eq:phi1:driftApprox}
\phi^{(1)}_\mu\lp E; y_{\rm cut}\rp \approx  \phi_\mu^{(0)}\mathcal{L}\int_{y_{\rm cut}}^1\dd y\,\frac{\dd\Gamma}{\dd y} \left[-1 + \lp1-y\rp^{A}\right]\,.
\eeq
\Cref{fig:hard-vs-soft} shows this estimate for several choices of $\gamma$, in comparison with the estimate of the contribution of diffusion in the soft régime, namely $d_{\mu}/b_{\mu}$, as a function of $y_{\rm cut}$ (see also \cref{fig:b_and_d}). The ratio $\phi^{(1)}_\mu\lp E; y_{\rm cut}\rp/\phi_\mu^{(0)}$ quantifies the fractional correction of hard scatterings to the soft volume.

Note that the hard correction is negative. This simply reflects the fact that, for a falling spectrum, a hard loss removes
more muons from a given energy bin than are replenished by muons injected from
higher energies. When this quantity becomes of order one, the perturbative expansion in hard
insertions breaks down. In that regime one should not interpret the negative
correction as a negative event rate or a negative volume; rather, the separation
between soft evolution and hard insertions is no longer under perturbative
control. Choosing a large enough value of $y_{\rm cut}$ makes the hard contribution a perturbative correction to diffusion in the soft regime. As can be seen, this setup does not allow predictions with a better precision than about $25\%$, which is, however, at the same level of the theoretical uncertainty on the QED rates (see \cref{sec:QEDlosses}) and more than enough to describe faithfully the observed data so far, at least for what concerns a diffuse neutrino flux. 

\bibliographystyle{JHEP}
\bibliography{biblio}

\end{document}